%% file: main.tex
\documentclass[sigconf]{acmart}

\usepackage{graphicx}
\usepackage{balance}  

\usepackage{graphics}
\usepackage[ruled, vlined, linesnumbered]{algorithm2e} 

\usepackage{booktabs}
\usepackage{soul}
\usepackage[center]{subfigure}

\usepackage{amsmath}
\usepackage{amsfonts}
\usepackage{amsbsy}
\usepackage{amsthm}
\usepackage[]{caption}
\usepackage{multirow}
\usepackage[mathscr]{eucal} 
\usepackage{tabularx}

\usepackage{bbding}

\let\oldnl\nl
\newcommand{\nonl}{\renewcommand{\nl}{\let\nl\oldnl}}

\DeclareMathOperator*{\argmin}{arg\,min}

\newcommand{\floor}[1]{\lfloor #1 \rfloor}
\usepackage{array}

\AtBeginDocument{%
  \providecommand\BibTeX{{%
    \normalfont B\kern-0.5em{\scshape i\kern-0.25em b}\kern-0.8em\TeX}}}

\copyrightyear{2021}
\acmYear{2021}
\setcopyright{iw3c2w3}
\acmConference[WWW '21]{Proceedings of the Web Conference 2021}{April 19--23, 2021}{Ljubljana, Slovenia}
\acmBooktitle{Proceedings of the Web Conference 2021 (WWW '21), April 19--23, 2021, Ljubljana, Slovenia}

\acmPrice{}
\acmDOI{10.1145/3442381.3450010}
\acmISBN{978-1-4503-8312-7/21/04}

\begin{document}
	
	\title{How Do Hyperedges Overlap in Real-World Hypergraphs? - Patterns, Measures, and Generators
	}
	
	\author{Geon Lee}
	\authornote{Equal Contribution.}
	\affiliation{%
		\institution{KAIST AI}
		\city{Daejeon}
		\country{South Korea}
	}
	\email{geonlee0325@kaist.ac.kr}
	
	\author{Minyoung Choe}
	\authornotemark[1]
	\affiliation{%
		\institution{KAIST AI}
		\city{Daejeon}
		\country{South Korea}
	}
	\email{minyoung.choe@kaist.ac.kr}
	
	\author{Kijung Shin}
	\affiliation{%
		\institution{KAIST AI \& EE}
		\city{Daejeon}
		\country{South Korea}
	}
	\email{kijungs@kaist.ac.kr}

	\begin{abstract}
		\input{000abstract.tex}

	\end{abstract}

\input{dfn.tex}
	\maketitle
	
	\section{Introduction}
	\label{sec:intro}
	\input{010intro.tex}
	
	\section{Related Work}
	\label{sec:related}
	\input{020related.tex}

	\section{Datasets and Null Models}
	\label{sec:prelim}
	\input{030prelim.tex}

	\section{Observations}
	\label{sec:observation}
	\input{040observation.tex}

	\section{Hypergraph Generation}
	\label{sec:method}
	\input{050method.tex}

	
	\section{Conclusions}
	\label{sec:summary}
	\input{070summary.tex}

	\vspace{1mm} 
	{\small \smallsection{Acknowledgements} We thank Dr. Jisu Kim for fruitful discussions.
		This work was supported by National Research Foundation of Korea (NRF) grant funded by the
		Korea government (MSIT) (No. NRF-2020R1C1C1008296) and Institute of Information \& Communications
		Technology Planning \& Evaluation (IITP) grant funded by the Korea government (MSIT) (No. 2019-0-00075, Artificial Intelligence Graduate School Program (KAIST)).}

	\appendix

	\bibliographystyle{ACM-Reference-Format}
	\bibliography{BIB/ref}

    \input{080appendix.tex}

\end{document}

%% file: 000abstract.tex
Hypergraphs, a generalization of graphs, naturally represent groupwise relationships among multiple individuals or objects, which are common in many application areas, including web, bioinformatics, and social networks.
The flexibility in the number of nodes in each hyperedge, which provides the expressiveness of hypergraphs, brings about structural differences between graphs and hypergraphs. 
Especially, the overlaps of hyperedges lead to complex high-order relations beyond pairwise relations, raising new questions that have not been considered in graphs: How do hyperedges overlap in real-world hypergraphs? Are there any pervasive characteristics? What underlying process can cause such patterns?

In this work, we closely investigate thirteen real-world hypergraphs from various domains and share interesting observations of the overlaps of hyperedges. To this end, we define principled measures and statistically compare the overlaps of hyperedges in real-world hypergraphs and those in null models. Additionally, based on the observations, we propose \method, a realistic hypergraph generative model. 
\method is \textbf{(a) Realistic:} it accurately reproduces overlapping patterns of real-world hypergraphs, \textbf{(b) Automatically Fittable:} its parameters can be tuned automatically using \methodauto to generate hypergraphs particularly similar to a given target hypergraph, \textbf{(c) Scalable:} it generates and fits a hypergraph with $0.7$ billion hyperedges within few hours.


%% file: dfn.tex
\newcommand\blue[1]{\textcolor{black}{#1}}
\newcommand\geon[1]{\textcolor{blue}{[Geon:#1]}}
\newcommand\minyoung[1]{\textcolor{brown}{[Minyoung:#1]}}
\newcommand\camera[1]{\textcolor{red}{#1}}

\newcommand{\smallsection}[1]{{\vspace{0.05in} \noindent {\bf{\underline{\smash{#1}}}}}}
\newtheorem{obs}{\textbf{Observation}}
\newtheorem{defn}{\textbf{Definition}}
\newtheorem{thm}{\textbf{Theorem}}
\newtheorem{axm}{\textbf{Axiom}}
\newtheorem{lma}{\textbf{Lemma}}

\newcommand{\cmark}{\ding{51}}%
\newcommand{\xmark}{\ding{55}}%

\newcommand{\method}{\textsc{HyperLap}\xspace}
\newcommand{\methodauto}{\textsc{HyperLap}\textsuperscript{+}\xspace}
\newcommand{\hypernull}{\textsc{HyperNULL}\xspace}
\newcommand{\hypercl}{\textsc{HyperCL}\xspace}
\newcommand{\hyperpa}{\textsc{HyperPA}\xspace}
\newcommand{\hyperff}{\textsc{HyperFF}\xspace}

\definecolor{myred}{RGB}{195, 79, 82}
\definecolor{mygreen}{RGB}{86, 167 104}
\definecolor{myblue}{RGB}{74, 113 175}

%% file: 010intro.tex
Group interactions among multiple individuals or objects are omnipresent in complex systems: collaborations of co-authors, co-purchases of items, group communications in question-and-answer sites, to name a few. They are naturally modeled as a \textit{hypergraph} where each \textit{hyperedge} (i.e., a subset of an arbitrary number of nodes) represents a group interaction. 
Hypergraphs are a generalization of ordinary graphs, which naturally describe pairwise interactions.


In real-world hypergraphs, hyperedges are overlapped with each other, revealing interesting relations between them.
Due to the flexibility in the size of each hyperedge, even a fixed number of hyperedges can overlap in infinitely many different ways.
Moreover, these relations are high-order, and decomposing them into pairwise relations loses considerable information.
This unique property of hypergraphs poses important  questions that have not been considered in graphs:
(1) How do hyperedges overlap in real-world hypergraphs?
(2) Are there any non-trivial patterns that distinguish real-world hypergraphs from random hypergraphs?
(3) How can we reproduce the patterns through simple mechanisms?



These questions are partially answered in recent empirical studies, which reveal structural and dynamical patterns of real-world hypergraphs.
The discovered patterns are regarding giant connected components \cite{do2020structural}, diameter \cite{do2020structural,kook2020evolution}, $3$-cliques~\cite{benson2018simplicial}, $3$-hyperedge subhypergraphs \cite{lee2020hypergraph}, simplicial closure~\cite{benson2018simplicial}, similarity between temporally close hyperedges \cite{benson2018sequences}, the number of intersecting hyperedges \cite{kook2020evolution}, etc.
These patterns are directly or indirectly affected by the overlaps of hyperedges.
Moreover, the overlaps of hyperedges have been considered for hyperedge prediction~\cite{benson2018sequences,benson2018simplicial,lee2020hypergraph} and realistic hypergraph generation~\cite{do2020structural}. 

In this work, we complement the previous studies with new findings, measures, and realistic generative models regarding the overlaps of hyperedges.
To this end, we closely examine thirteen real-world hypergraphs from six distinct domains.
Specifically, we analyze the overlaps of hyperedges in them at three different levels: subsets of nodes, hyperedges, and egonets. Then, we verify our findings using randomized hypergraphs, where we overlap hyperedges randomly while preserving the degrees of nodes and the sizes of hyperedges.
Our investigation reveals that the overlaps of hyperedges in real-world hypergraphs show the following properties:

\setlength{\leftmargini}{1em}
\begin{itemize}
	\item \textbf{Substantial:} Hyperedges in each egonet tend to overlap more substantially in real-world hypergraphs than in randomized ones. 
	\item \textbf{Heavy-tailed:} The number of hyperedges overlapping at each pair or triple of nodes is more skewed with a heavier tail in real-world hypergraphs than in randomized ones. The number of overlapping hyperedges follows a near power-law distribution.
	\item \textbf{Homophilic:} Nodes contained in each hyperedge tend to be structurally more similar (i.e., more hyperedges overlap at them) in real-world hypergraphs than in randomized ones.
\end{itemize}

\begin{table*}[t!]
\begin{center}
\vspace{-3mm}
\caption{\label{tab:crown}\methodauto accurately reproduces the overlaps of hyperedges in real-world hypergraphs. Synthetic hypergraphs created by \methodauto exhibit (Obs.~\ref{obs:density}) dense egonets, (Obs.~\ref{obs:overlapness}) highly overlapped egonets, (Obs.~\ref{obs:pair}) heavy-tailed pair-of-nodes degree distribution, (Obs.~\ref{obs:triple}) heavy-tailed triple-of-nodes degree distribution, and (Obs.~\ref{obs:hyperedge_locality}) homogeneous hyperedges. We provide the full results in ~\cite{online2020appendix}.}
\scalebox{1.00}{
\begin{tabular}{c|ccccc}
 \toprule
 & Observation~\ref{obs:density} & Observation~\ref{obs:overlapness} & Observation~\ref{obs:pair} & Observation~\ref{obs:triple} & Observation~\ref{obs:hyperedge_locality}\\
 \hline
 \multirow{6}{*}{\shortstack{Real \\(threads-math)}} &
 \raisebox{-\totalheight}{\includegraphics[width=0.145\textwidth]{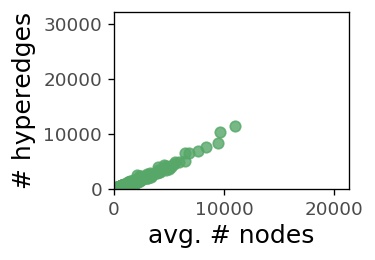}} & \raisebox{-\totalheight}{\includegraphics[width=0.145\textwidth,]{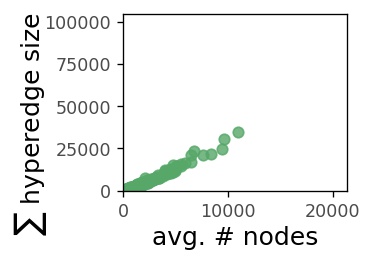}} & \raisebox{-\totalheight}{\includegraphics[width=0.145\textwidth]{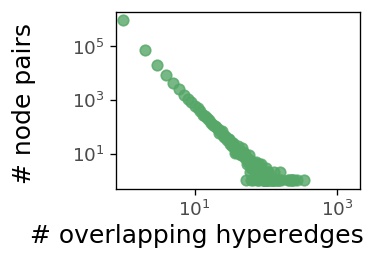}} & \raisebox{-\totalheight}{\includegraphics[width=0.145\textwidth]{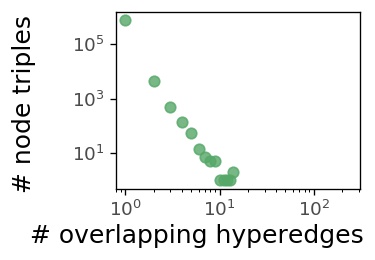}} & \raisebox{-\totalheight}{\includegraphics[width=0.145\textwidth]{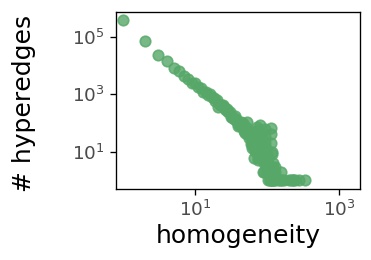}} \\
 \hline
 \multirow{6}{*}{\textbf{\shortstack{\methodauto\\(Proposed)}}} &
 \raisebox{-\totalheight}{\includegraphics[width=0.145\textwidth]{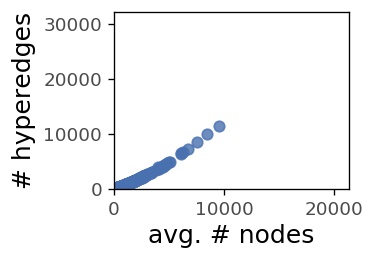}} &  \raisebox{-\totalheight}{\includegraphics[width=0.145\textwidth,]{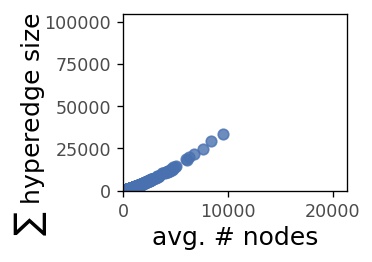}} &  \raisebox{-\totalheight}{\includegraphics[width=0.145\textwidth]{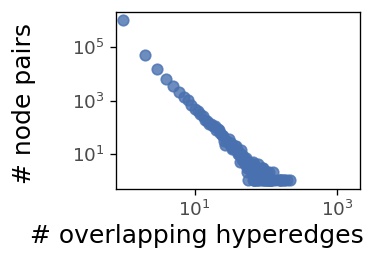}} &  \raisebox{-\totalheight}{\includegraphics[width=0.145\textwidth]{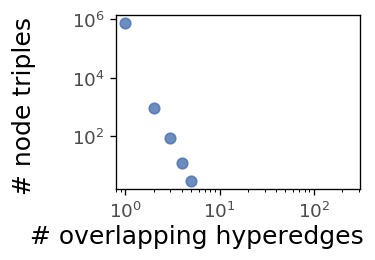}} &  \raisebox{-\totalheight}{\includegraphics[width=0.145\textwidth]{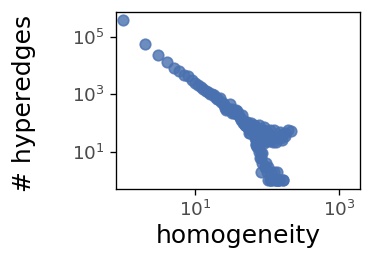}} 
 \\
 \hline
 \multirow{6}{*}{\shortstack{\shortstack{\hyperpa\\(Competitor)}}} &
 \raisebox{-\totalheight}{\includegraphics[width=0.145\textwidth]{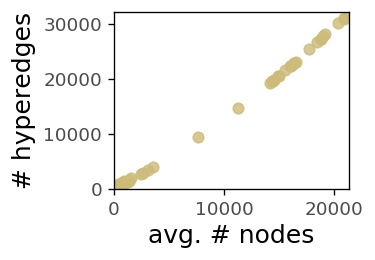}} &  \raisebox{-\totalheight}{\includegraphics[width=0.145\textwidth,]{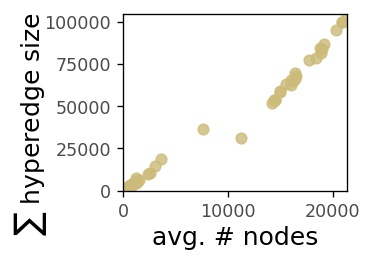}} &  \raisebox{-\totalheight}{\includegraphics[width=0.145\textwidth]{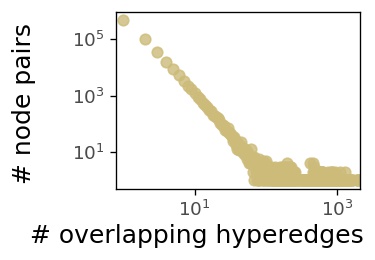}} &  \raisebox{-\totalheight}{\includegraphics[width=0.145\textwidth]{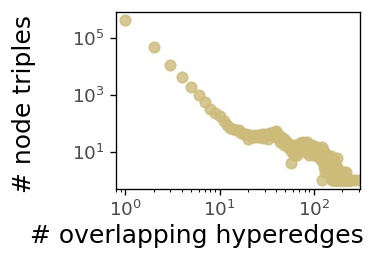}} &  \raisebox{-\totalheight}{\includegraphics[width=0.145\textwidth]{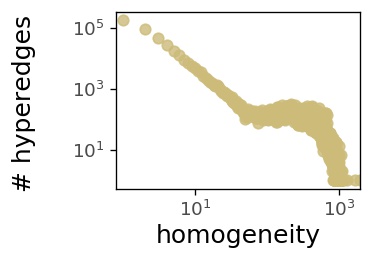}} \\
 \bottomrule
 \end{tabular}}
\end{center}
\end{table*}

For the investigation of real-world hypergraphs, we design novel and principled measures.
We show that our measure of overlapness of hyperedges satisfies three intuitively clear axioms, while a widely-used density measure does not.
We also introduce a measure of overlapness at subsets of nodes, which reveals interesting near power-law behaviors, and
a measure of homogeneity of hyperedges, which plays a key role in realistic hypergraph generation.


What underlying process can cause hyperedges to systematically overlap exhibiting the above patterns?
We design \method, a stochastic hypergraph generative model. 
\method accurately reproduces realistic overlapping patterns of hyperedges.
In addition, we present \methodauto, which automatically tune\blue{s} the parameters of \method to generate synthetic hypergraphs particularly similar to a given target graph (see Table~\ref{tab:crown}).
\method gives intuitions useful in reasoning about and predicting the evolution of the hypergraphs, and it can be used to generate synthetic hypergraphs for simulations and evaluation of algorithms when it is impossible to collect or track real hypergraphs. 
\methodauto can be used to anonymize hypergraphs that cannot be publicized to share them. 


Our contributions are summarized as follow:
\setlength{\leftmargini}{1em}
\begin{itemize}
    \item \textbf{Observations in Real-world Hypergraphs:} We discover three unique characteristics of the overlaps of hyperedges in real-world hypergraphs, and we verify them using randomized hypergraphs.
	\item \textbf{Novel Measures:} We define novel and principled measures regarding the overlaps of hyperedges at $3$ different levels. They play key roles in investigation and realistic hypergraph generation.
	\item \textbf{Realistic Generative Model:} We propose \method, a stochastic hypergraph generator that reproduces realistic overlaps of hyperedges. We also provide \methodauto, which automatically fit\blue{s} the parameters of \method to a given hypergraph.
	Empirically, they scale near linearly with the number of hyperedges.
\end{itemize}

\noindent \textbf{Reproducibility:}  The source code and datasets used in this work are available at \blue{\url{https://github.com/young917/www21-hyperlap}}.

In Section~\ref{sec:related}, we discuss related work. In Section~\ref{sec:prelim}, we describe the datasets and the null models used throughout this work. In Section~\ref{sec:observation}, we share our observations of the overlaps of hyperedges in real-world hypergraphs. In Section~\ref{sec:method}, we propose \method, a realistic hypergraph generative model, and provide experimental results. 
Lastly, we offer conclusions in Section~\ref{sec:summary}.

%% file: 020related.tex
There have been extensive studies on macroscopic structural patterns \cite{barabasi1999emergence,watts1998collective,faloutsos1999power,shin2018patterns}, microscopic structural patterns \cite{milo2004superfamilies,milo2002network}, and dynamical patterns \cite{shin2017wrs,hidalgo2008dynamics,leskovec2007graph} in real-world pairwise graphs, and numerous realistic graph generators \cite{leskovec2007graph,leskovec2007scalable,you2018graphrnn,goyal2020graphgen,chakrabarti2004r} for reproducing the discovered patterns have been proposed.
In this section, we focus on hypergraphs and review previous studies on empirical patterns in real-world hypergraphs and realistic hypergraph generators. 
Hypergraphs have been used in a wide range of fields, including computer vision~\cite{yu2012adaptive}, bioinformatics~\cite{hwang2008learning}, circuit design~\cite{karypis1999multilevel}, social network analysis~\cite{yang2019revisiting}, and recommendation~\cite{mao2019multiobjective}. They have been used in various analytical and learning tasks, including classification~\cite{jiang2019dynamic,yadati2019hypergcn}, clustering~\cite{amburg2020clustering,li2017inhomogeneous,li2018submodular}, and hyperedge prediction~\cite{benson2018simplicial,yoon2020much}.
In addition to the realistic hypergraph generators described below, a number of random hypergraph models \cite{stasi2014beta,karonski2002phase,berge1989hyper,chodrow2020configuration} have been used for statistical tests.

Benson et al. \cite{benson2018simplicial} focused on simplicial closure events (i.e., the first appearance of a hyperedge containing a set of nodes each of whose pairs co-appear in previous hyperedges) and investigated how their probabilities are affected by local features, such as average degree, in real-world hypergraphs from different domains.

Benson et al. \cite{benson2018sequences} considered sequences (i.e., time-ordered hyperedges that are relevant to each other) in real-world hypergraphs and showed that hypergraphs in a sequence tend to be more similar to recent hyperedges than distant ones. They also discovered that the number of hyperedges overlapping at each pair and triple of nodes tends to be larger in each sequence than in a null model. 
In addition, the authors proposed to exploit both patterns when predicting the next hyperedge in a sequence.
Notably, in Section~\ref{sec:observation:pair_triple_level}, we also examine the number of hyperedges overlapping at each pair and triple of nodes. However, we (a) examine them at the hypergraph level, (b) discover their near power-law distributions, and (3) compare them with those in degree-preserving randomized hypergraphs.

Do et al. \cite{do2020structural} considered projecting a real-world hypergraph into multiple pairwise graphs so that each $k$-th graph describes the interactions between size-$k$ subsets of nodes. They showed that the pairwise graphs exhibit (a) heavy-tailed degree and singular-value distributions, (b) giant connected components, (c) small diameter, and (d) high clustering coefficients.
Inspired by the observations, the authors proposed a hypergraph generator called \hyperpa~\cite{do2020structural}.
In \hyperpa, the subset of nodes that form a hyperedge with a new node is selected with probability proportional to the number of hyperedges containing the subset.

Kook et al. \cite{kook2020evolution} revealed that the ratio of intersecting hyperedges and the diameter of real-world hypergraph decrease\blue{s} over time, while the number of hyperedges increases faster than the number of nodes.
Additionally, they discovered four structural patterns regarding (a) the number of hyperedges containing each node, (b) the size of hyperedges, (c) the size of intersections between two hyperedges, and (d) singular values of incident matrices.
In order to reproduce the patterns, the authors proposed a hypergraph generator called \hyperff.
For each new node, \hyperff simulates forest fire spreading over hyperedges, and the new node forms a size-$2$ hyperedge with each burned node. Then, \hyperff simulates forest fire again to expand each size-$2$ hyperedge.

Lee et al. \cite{lee2020hypergraph} proposed 26 hypergraph motifs (h-motifs), which are connectivity patterns of three connected hyperedges, based on the emptiness of the seven Venn diagram regions. They showed that the relative occurrences of the h-motifs are particularly similar in real-world hypergraphs from the same domain.

All these findings are directly or indirectly related to the overlaps of hyperedges. In this work, we complement the previous studies with new findings, measures, and more realistic and scalable generators, all of which are related to the overlaps of hyperedges.

%% file: 030prelim.tex
\begin{table}[t!]
	\begin{center}
		\caption{\label{notations}Frequently-used symbols.}
		\scalebox{0.925}{
			\begin{tabular}{c|l}
				\toprule
				\textbf{Notation} & \textbf{Definition}\\
				\midrule
				$G=(V,E)$ & hypergraph with nodes $V$ and hyperedges $E$\\
				$E=\{e_1,...,e_{|E|}\}$ & set of hyperedges\\
				$E_{\{v\}}$ & set of hyperedges that contain a node $v$\\
				$E_{S}$ & set of hyperedges that contain a subset $S$ of nodes\\
				\midrule
				$L$ & number of levels in \method\\
				$w_1,...,w_L$ & weight of each level\\
				$S_{g}^{(\ell)}$ & set of nodes in a group \textsl{g} of level $\ell$\\
				\bottomrule 
			\end{tabular}}
	\end{center}
\end{table}

In this section, we first introduce some notations and preliminaries. Then, we describe the datasets and the null models used throughout this paper. Refer to Table~\ref{notations} for the frequently-used notations.

\subsection{Preliminaries and Notations}
\label{sec:prelim:preliminaries}
We review the concept of hypergraphs and then the Chung-Lu model, which our null model is based on.

\smallsection{Hypergraphs:}
A \textit{hypergraph} $G=(V,E)$ consists of a set of nodes $V$ and a set of hyperedges $E\subseteq2^{V}$. Each hyperedge $e\subseteq V$ is a non-empty subset of $|e|$ nodes. 
For each node $v$, we denote the set of hyperedges that contain $v$ by $E_{\{v\}}:=\{e\in E: v\in e\}$, and
the degree $d_{v}:=|E_{\{v\}}|$ of $v$ is defined as the number of hyperedges that contains $v$.
We say two hyperedges $e_i$ and $e_j$ are \textit{overlapped} or \textit{intersected} if they share any node, i.e., $e_i \cap e_j \neq \varnothing$. 

\smallsection{Chung-Lu Models:}
The Chung-Lu (CL) model~\cite{chung2002average} is a random graph model, and it yields graphs where a given degree sequence of nodes is expected to be preserved.
Consider a graph $\bar{G}=(\bar{V},\bar{E})$ where $\bar{E}$ is a set of pairwise edges. Given a desired degree distribution $\{d_1,d_2,...,d_{|\bar{V}|}\}$, where $d_i$ is the degree of the node $i$, the CL model generates a random graph by creating an edge between each pair of nodes with probability proportional to the product of their degrees.
That is, for each pair $(i,j)$ of nodes, the edge $e_{ij}$ is created with probability $\frac{d_i d_j}{2M}$, where $M=\frac{1}{2}\sum_{k=1}^{|\bar{V}|} d_k$, assuming $d_k<\sqrt{M}$ holds for all $k$.
If we let $\tilde{d_i}$ be the degree of each node $i$ in the generated graph, its expected value is equal to $d_i$, i.e.,
\begin{equation*}
    \mathbb{E}[\tilde{d_i}]=\sum\nolimits_{j=1}^{|\bar{V}|} \frac{d_i d_j}{2M} = d_i \sum\nolimits_{j=1}^{|\bar{V}|} \frac{d_j}{2M} = d_i.
\end{equation*}

While the CL model flips a coin for all possible $O(|\bar{V}|^2)$ node pairs, the fast CL (FCL) model~\cite{pinar2012similarity} samples two nodes independently with probability proportional to the degree of each node. 
Then, it creates an edge between the sampled pair of nodes.
This process is repeated $|\bar{E}|$ times, and the total time complexity is $O(|\bar{E}|)$.
Even in graphs generated by the FCL model, the expected degree of each node $i$ is equal to $d_i$.

\begin{table}[t!]
	\begin{center}
		\caption{\label{tab:datasets}Summary statistics of 13 real-world hypergraphs from 6 domains: the number of nodes $|V|$, the number of hyperedges $|E|$, the average hyperedge size $\mathrm{avg}_{e\in E}|e|$, and the maximum hyperedge size $\max_{e\in E}|e|$.}
		\scalebox{0.925}{
			\begin{tabular}{l|c|c|c|c}
				\toprule
				\textbf{Dataset} & $\mathbf{|V|}$ & $\mathbf{|E|}$ & $\mathbf{\mathrm{avg}_{e\in E}|e|}$ & $\mathbf{\max_{e\in E}|e|}$\\
				\midrule
				email-Enron & 143 & 1,459 & 3.13 & 37 \\
				email-Eu & 986 & 24,520 & 3.62 & 40 \\
				\midrule
				contact-primary & 242 & 12,704 & 2.41& 5 \\
				contact-high & 327 & 7,818 & 2.32 & 5 \\
				\midrule
				NDC-classes & 1,149 & 1,049 & 6.16 & 39 \\
				NDC-substances & 3,767 & 6,631 & 9.70 & 187 \\
				\midrule
				tags-ubuntu & 3,021 & 145,053 & 3.42 & 5 \\
				tags-math & 1,627 & 169,259 & 3.49 & 5 \\
				\midrule
				threads-ubuntu & 90,054 & 115,987 & 2.30 & 14 \\
				threads-math & 153,806 & 535,323 & 2.61 & 21 \\
				\midrule
				coauth-DBLP & 1,836,596 & 2,170,260 & 3.43 & 280 \\
				coauth-geology & 1,091,979 & 909,325 & 3.87 & 284 \\
				coauth-history & 503,868 & 252,706 & 3.01 & 925 \\
				\bottomrule 
			\end{tabular}}
	\end{center}
\end{table}

\subsection{Datasets~\label{sec:prelim:datasets}}
We use thirteen real-world hypergraphs from six different domains~\cite{benson2018simplicial} 
after removing duplicated or singleton hyperedges.
Refer to Table~\ref{tab:datasets} for some statistics of the hypergraphs.

\setlength{\leftmargini}{1em}
\begin{itemize}
    \item \textbf{email} (email-Enron~\cite{klimt2004enron} and email-Eu~\cite{leskovec2005graphs,yin2017local}): Each node is an email account and each hyperedge is a set of the sender and receivers of an email.
    \item \textbf{contact} (contact-primary~\cite{stehle2011high} and contact-high~\cite{mastrandrea2015contact}): Each node is a person, and each hyperedge is a group interaction among individuals.
    \item \textbf{drugs} (NDC-classes and NDC-substances): Each node is a class label (in NDC-classes) or a substances (in NDC-substances) and each hyperedge is a set of labels/substances of a drug.
    \item \textbf{tags} (tags-ubuntu and tags-math): Each node is a tag, and each hyperedge is a set of tags attached to a question.
    \item \textbf{threads} (threads-ubuntu and threads-math): Each node is a user, and each hyperedge is a group of users participating in a thread.
    \item \textbf{co-authorship} (coauth-DBLP, coauth-geology~\cite{sinha2015overview}, and coauth-history~\cite{sinha2015overview}): Each node is an author and each hyperedge is a set of authors of a publication.
\end{itemize}

\subsection{Null Model: \hypercl (Algorithm~\ref{alg:random_hypergraph})~\label{sec:prelim:random_hypergraphs}}
We introduce \hypercl, a random hypergraph generator that extends the FCL model (see Section~\ref{sec:prelim:preliminaries}) to hypergraphs. We use random hypergraphs generated by \hypercl as null models throughout this work.
As described in Algorithm~\ref{alg:random_hypergraph}, the degree distribution of nodes and the size distribution of hyperedges in a considered real-world hypergraph are given as inputs. 
For each $i$-th hyperedge $\tilde{e}_i$, its nodes are sampled independently, with probability proportional to the degree of each node (i.e., the probability is $d_v/\sum_{j=1}^{|V|}d_j$ for each node $v$) until the size of the hyperedge reaches $s_i$ (lines~\ref{alg:random_hypergraph:size}-\ref{alg:random_hypergraph:include}). Note that duplicated nodes are ignored so that each $i$-th hypergraph contains $s_i$ distinct nodes. 

In hypergraphs generated by \hypercl, the size distribution of hyperedges is exactly the same as the input size distribution, and the degree distribution of nodes is also expected to be similar to the input degree distribution.
Specifically, if we assume $\sum_{j=1}^{|V|} d_j \gg (\max_{k\in  \{1,\cdots,|E|\}}s_{k}) \cdot (\max_{k\in  \{1,\cdots,|V|\}}d_{k}$) and let $\tilde{d_v}$ be the in a generated hypergraph,
\begin{align*}
    \mathbb{E}[\tilde{d_v}] &  =  \sum\nolimits_{\tilde{e}\in E}P[v\in \tilde{e}] \\
    & \approx \sum\nolimits_{\tilde{e}\in E}\left(|\tilde{e}|\cdot \frac{d_v}{\sum_{j=1}^{|V|} d_j}\right)=\frac{d_v}{\sum_{j=1}^{|V|} d_j}\sum\nolimits_{\tilde{e}\in E}|\tilde{e}|=d_v.
\end{align*}





\noindent We show experimentally in \cite{online2020appendix} that the degree distributions in hypergraphs generated by \hypercl are closed to the input degree distribution.

%% file: 040observation.tex
\begin{table*}[t!]
\begin{center}
\caption{\label{tab:observation:summary}Hyperedges in \textcolor{mygreen}{real-world hypergraphs} overlap distinctly from those in \textcolor{myred}{randomized hypergraphs}. We examine (Obs. 1) density of each egonet, (Obs. 2) overlapnesses of each egonet, (Obs. 3) the number of hyperedges overlapping at each pair of nodes, (Obs. 4) the number of hyperedges overlapping at each triple of nodes, and (Obs. 5)  homogeneity of each hyperedge. \blue{Regarding Observation 5, we preprocessed the continuous values of hyperedge homogeneity by binning them into their nearest integers.} We provide the full results in ~\cite{online2020appendix}.}
\scalebox{0.95}{
\begin{tabular}{c|cccccc}
 \toprule
 & \textbf{email-Eu} & \textbf{contact-primary} & \textbf{NDC-substances} & \textbf{tags-math} & \textbf{threads-ubuntu} & \textbf{coauth-DBLP}\\ 
\hline
\parbox[t]{2mm}{\multirow{6}{*}{\rotatebox[origin=c]{90}{\ \ Observation~\ref{obs:density}}}} & \raisebox{-\totalheight}{\includegraphics[width=0.1475\textwidth]{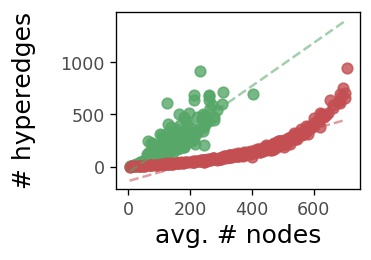}} & \raisebox{-\totalheight}{\includegraphics[width=0.1475\textwidth,]{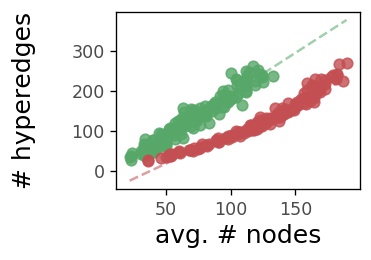}} & \raisebox{-\totalheight}{\includegraphics[width=0.1475\textwidth]{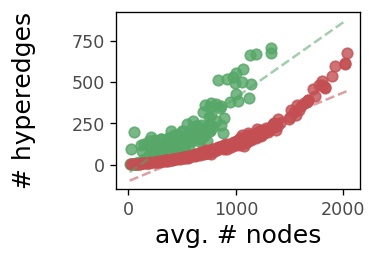}} & \raisebox{-\totalheight}{\includegraphics[width=0.1475\textwidth]{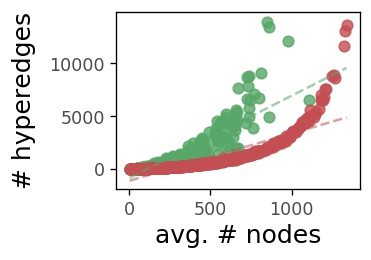}} & \raisebox{-\totalheight}{\includegraphics[width=0.1475\textwidth]{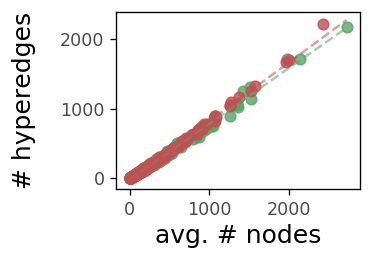}} & \raisebox{-\totalheight}{\includegraphics[width=0.1475\textwidth]{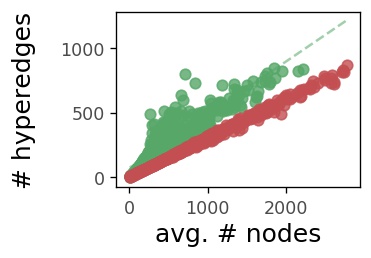}}\\
\hline
\parbox[t]{2mm}{\multirow{6}{*}{\rotatebox[origin=c]{90}{\ Observation~\ref{obs:overlapness}}}} & \raisebox{-\totalheight}{\includegraphics[width=0.1475\textwidth]{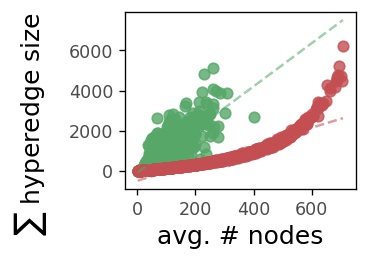}} & \raisebox{-\totalheight}{\includegraphics[width=0.1475\textwidth,]{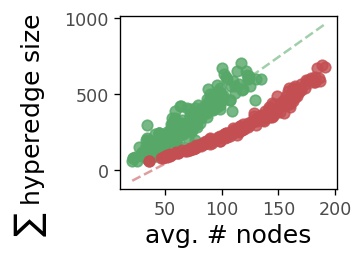}} & \raisebox{-\totalheight}{\includegraphics[width=0.1475\textwidth]{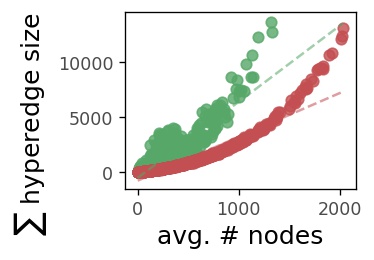}} & \raisebox{-\totalheight}{\includegraphics[width=0.1475\textwidth]{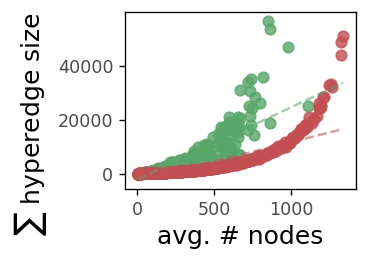}} & \raisebox{-\totalheight}{\includegraphics[width=0.1475\textwidth]{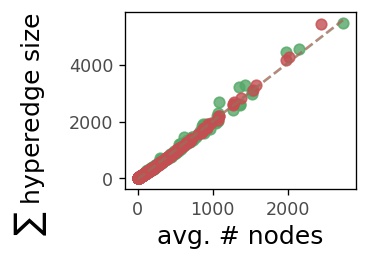}} & \raisebox{-\totalheight}{\includegraphics[width=0.1475\textwidth]{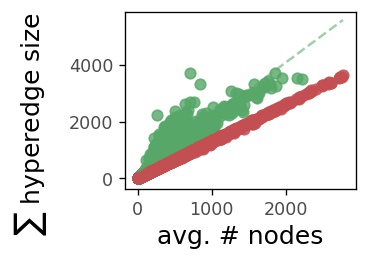}}\\
 \hline 
\parbox[t]{2mm}{\multirow{6}{*}{\rotatebox[origin=c]{90}{\ \ \ Observation~\ref{obs:pair}}}} & \raisebox{-\totalheight}{\includegraphics[width=0.1475\textwidth]{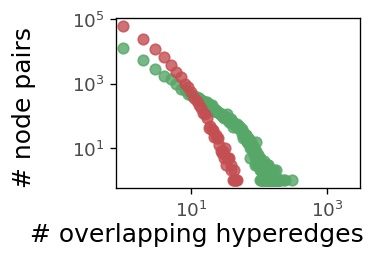}} & \raisebox{-\totalheight}{\includegraphics[width=0.1475\textwidth,]{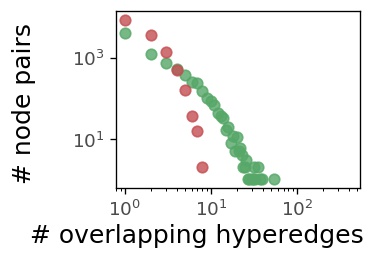}} & \raisebox{-\totalheight}{\includegraphics[width=0.1475\textwidth]{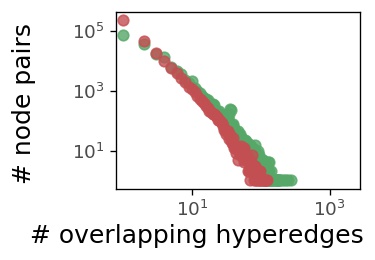}} & \raisebox{-\totalheight}{\includegraphics[width=0.1475\textwidth]{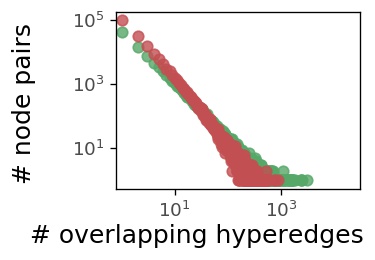}} & \raisebox{-\totalheight}{\includegraphics[width=0.1475\textwidth]{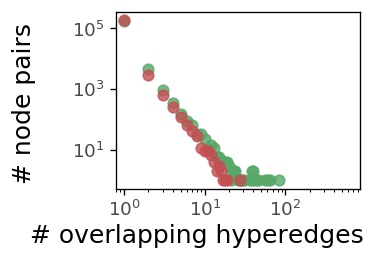}} & \raisebox{-\totalheight}{\includegraphics[width=0.1475\textwidth]{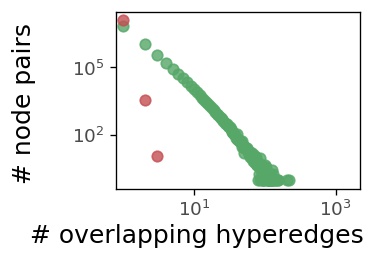}}\\
\hline
\parbox[t]{2mm}{\multirow{6}{*}{\rotatebox[origin=c]{90}{\ \ Observation~\ref{obs:triple}}}} & \raisebox{-\totalheight}{\includegraphics[width=0.1475\textwidth]{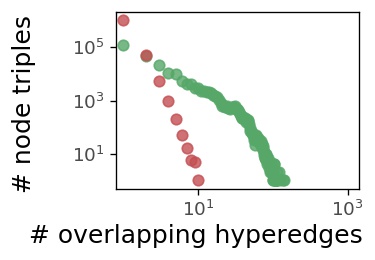}} & \raisebox{-\totalheight}{\includegraphics[width=0.1475\textwidth,]{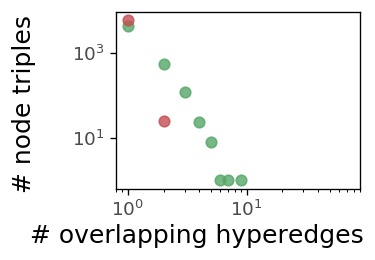}} & \raisebox{-\totalheight}{\includegraphics[width=0.1475\textwidth]{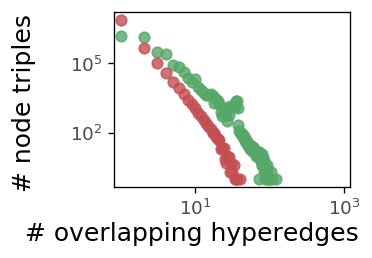}} & \raisebox{-\totalheight}{\includegraphics[width=0.1475\textwidth]{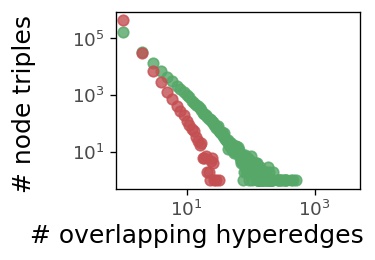}} & \raisebox{-\totalheight}{\includegraphics[width=0.1475\textwidth]{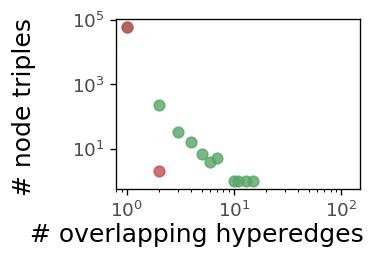}} & \raisebox{-\totalheight}{\includegraphics[width=0.1475\textwidth]{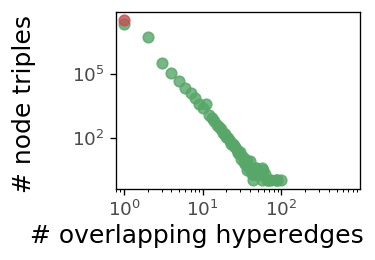}}\\
\hline
\parbox[t]{2mm}{\multirow{6}{*}{\rotatebox[origin=c]{90}{\ \ Observation~\ref{obs:hyperedge_locality}}}} & \raisebox{-\totalheight}{\includegraphics[width=0.1475\textwidth]{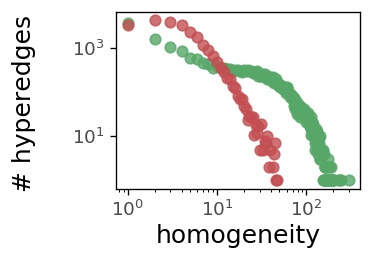}} & \raisebox{-\totalheight}{\includegraphics[width=0.1475\textwidth,]{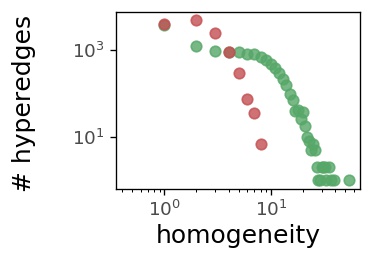}} & \raisebox{-\totalheight}{\includegraphics[width=0.1475\textwidth]{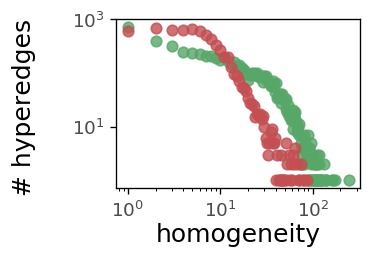}} & \raisebox{-\totalheight}{\includegraphics[width=0.1475\textwidth]{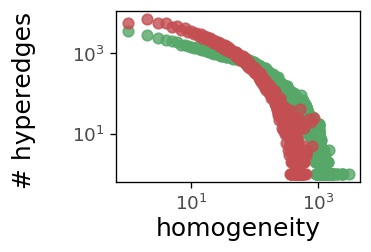}} & \raisebox{-\totalheight}{\includegraphics[width=0.1475\textwidth]{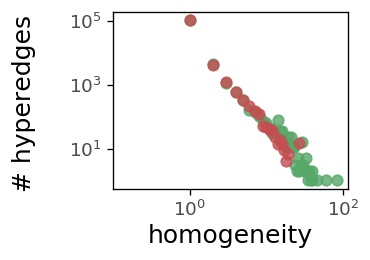}} & \raisebox{-\totalheight}{\includegraphics[width=0.1475\textwidth]{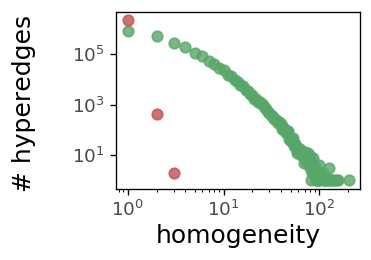}}\\
 \bottomrule
 \end{tabular}}
\end{center}
\end{table*}

In this section, we examine overlapping patterns of hyperedges in real-world hypergraphs, and we verify them by comparison with those in randomized hypergraphs obtained by \hypercl.
We investigate the overlaps of hyperedges at three different levels, and our observations are summarized as follow.

\begin{algorithm}[t]
	\caption{\hypercl: Random Hypergraph Generator \label{alg:random_hypergraph}}
	\SetKwInOut{Input}{Input}
	\SetKwInOut{Output}{Output}
	\Input{ (1) distribution of hyperedge sizes $\{s_1,...,s_{|E|}\}$\\ (2) distribution of node degrees $\{d_1,...,d_{|V|}\}$}
	\Output{ random hypergraph $\tilde{G}=(\tilde{V},\tilde{E})$}
	$\tilde{V}\leftarrow V$ and
	$\tilde{E}\leftarrow \varnothing $\\
	\For{\upshape\textbf{each } $i=1,...,|E|$}{
		$\tilde{e_i}\leftarrow \varnothing$\\
		\While{$|\tilde{e_i}| < s_i$ \label{alg:random_hypergraph:size}}{
			$v \leftarrow$ select a node with prob. proportional to the degree\label{alg:random_hypergraph:node_biased} \\
			$\tilde{e_i}\leftarrow\tilde{e_i}\cup \{v\}$\label{alg:random_hypergraph:include}\\
		} 
		$\tilde{E}\leftarrow\tilde{E}\cup\{\tilde{e_i}\}$\\
	}
	\Return{$\tilde{G}=(\tilde{V},\tilde{E})$}
\end{algorithm}

\setlength{\leftmargini}{1em}
\begin{itemize}
    \item \textbf{(L1) Egonet Level:} The overlaps of hyperedges in the egonet of each node tend to be more substantial in real-world hypergraphs than in randomized ones. 
    \item \textbf{(L2) Pair/Triple of Nodes Level:} 
    The number of hyperedges overlapping at each pair or triple of nodes follows a near (truncated) power-law distribution.
    Moreover, the number of overlapping hyperedges is more skewed with a heavier tail in real-world hypergraphs than in randomized ones.
    \item \textbf{(L3) Hyperedge Level:} Hyperedges tend to contain nodes that are structurally more similar (i.e., nodes where more hyperedges overlap) in real-world hypergraphs than in randomized ones. 
\end{itemize}

\subsection{L1. Egonet Level}~\label{sec:observation:egonet_level}

\smallsection{Density of Egonets:}
We first investigate egonets in real-world hypergraphs.
We define the egonet of a node $v$ as the set of hyperedges that contains $v$ (i.e., $E_{\{v\}}:=\{e\in E: v\in e\}$).
To quantitatively measure how substantially the hyperedges in an egonet overlap each other, we first consider the density (see Definition~\ref{defn:density}) of the egonets in real-world and randomized hypergraphs, and this leads to Observation~\ref{obs:density}.
\blue{While one might expect the density of a set of hyperedges $\mathcal{E}$ to be defined as the number of hyperedges divided by the size of the powerset of the induced nodes $\mathcal{V}$ (i.e., $\frac{|\mathcal{E}|}{2^{|\mathcal{V}|}-1}$), we follow the definition in \cite{hu2017maintaining} in this work.}

\begin{defn}[Density \cite{hu2017maintaining}] \label{defn:density}
    Given a set of hyperedges $\mathcal{E}$, the density of the set, $\rho(\mathcal{E})$ is defined as:
    \begin{equation*}
        \rho(\mathcal{E}) := \frac{|\mathcal{E}|}{|\bigcup_{e \in \mathcal{E}} e|}.
    \end{equation*}
\end{defn}

\begin{obs}\label{obs:density}
	Egonets in real-world hypergraphs tend to be denser than those in randomized hypergraphs. 
\end{obs}

Specifically, as seen in the figures in the first row of Table~\ref{tab:observation:summary}, 
when considering the egonets with the same number of hyperedges, they tend to contain fewer nodes in real-world hypergraphs than in randomized ones.
Thus, the density, which is defined as the ratio of the number of hyperedges to the number of nodes tends to be higher in real-world hypergraphs than in randomized ones.
In the figures, the slopes of the regression lines, which are close to the average egonet density, are steeper in real-world hypergraphs than in randomized ones.

 
\smallsection{Principled Measure: Overlapness:}
However, density does not fully take the overlaps of hyperedges into consideration. Consider two sets of hyperedges: $\mathcal{E}_1=\{\{a,b,c\},\{a,b,c,d\},\{a,b,c,d,e\}\}$ and $\mathcal{E}_2=\{\{v,w,x\},\{x,y\},\{y,z\}\}$. 
While, intuitively, $\mathcal{E}_1$ are overlapped more substantially than $\mathcal{E}_2$,
the densities of both sets, which consist of the same numbers of nodes and hyperedges, are the same.

To address this issue, we first present three axioms that any reasonable measure of the hyperedge overlaps should satisfy. Then, we propose \textit{overlapness}, a new measure that satisfies all the axioms.
The three axioms are formalized in Axioms~\ref{axm:num_hyperedges}, \ref{axm:num_nodes}, and \ref{axm:size_hyperedges}.

\begin{axm}[Number of Hyperedges\label{axm:num_hyperedges}]
    Consider two sets of hyperedges $\mathcal{E}$ and $\mathcal{E}'$ that  contain hyperedges of the same size, and the same number of distinct nodes. Then, the set with more hyperedges is more overlapped  than the other.
    Formally, 
\begin{align*}
        \Big((|\mathcal{E}|<|\mathcal{E}'|)& \wedge (|e|=|e'|=n, \forall e\in \mathcal{E}, \forall e'\in \mathcal{E'}) 
         \\ & \wedge (|\bigcup\limits_{e\in \mathcal{E}}e|=|\bigcup\limits_{e'\in \mathcal{E'}}e'|)\Big)  \Rightarrow f(\mathcal{E}) < f(\mathcal{E}').
\end{align*}
\end{axm}

\begin{axm}[Number of Distinct Nodes\label{axm:num_nodes}]
    Consider two hyperedges $\mathcal{E}=\{e_1,\cdots,e_{n}\}$ and $\mathcal{E}'=\{e_1^{'},\cdots,e_{n}^{'}\}$ with the same number of hyperedges and the same size distribution of hyperedges.
    Then, the set containing less distinct nodes is more overlapped  than the other. Formally,
    \begin{align*}
        \Big((|\mathcal{E}|=|\mathcal{E}'|=n) & \wedge (|e_i|=|e'_i|, \forall i\in \{1,\cdots,n\})  \\ & \wedge (|\bigcup\limits_{e\in \mathcal{E}}e| > |\bigcup\limits_{e'\in \mathcal{E'}}e'|)  \Big)
        \Rightarrow f(\mathcal{E}) < f(\mathcal{E}').
    \end{align*}
\end{axm}


\begin{axm}[Sizes of Hyperedges\label{axm:size_hyperedges}]
   Consider two sets of hyperedges $\mathcal{E}=\{e_1,\cdots,e_{n}\}$ and $\mathcal{E}'=\{e_1^{'},\cdots,e_{n}^{'}\}$ with the same number of distinct nodes and the same number of hyperedges.
   Then, the set with larger hyperedges is more overlapped than the other. Formally,
    \begin{align*}
        \Big((|\mathcal{E}|=|\mathcal{E}'|=n) & \wedge (|e_i|<|e'_i|)  \wedge (|e_j|\leq|e'_j|, \forall j\in \{1,\cdots,n\}\setminus \{i\}) \\ & \wedge  (|\bigcup\limits_{e\in \mathcal{E}}e| = |\bigcup\limits_{e'\in \mathcal{E'}}e'|)  \Big)
        \Rightarrow f(\mathcal{E}) < f(\mathcal{E}').
    \end{align*}
\end{axm}




\begin{table}[t!]
	\begin{center}
		\caption{\label{tab:axioms}
		Overlapness measures the degree of hypergraph overlaps reasonably, satisfying all the axioms, while the others do not. 
		See Appendix~\ref{sec:appendix:overlapness_axiom} for details.}
		\scalebox{0.90}{
			\begin{tabular}{l|c|c|c}
				\toprule
				\textbf{Metric} & \textbf{Axiom 1} & \textbf{Axiom 2} & \textbf{Axiom 3}\\
				\midrule
				Intersection & \XSolidBrush & \XSolidBrush & \XSolidBrush\\
				Union Inverse & \XSolidBrush & \Checkmark & \XSolidBrush\\
				Jaccard Index & \XSolidBrush & \XSolidBrush & \XSolidBrush \\
				Overlap Coefficient & \XSolidBrush & \XSolidBrush & \XSolidBrush\\
				Density & \Checkmark & \Checkmark & \XSolidBrush \\
				\midrule
				\textbf{Overlapness} (Proposed) & \Checkmark & \Checkmark & \Checkmark\\
				\bottomrule 
			\end{tabular}}
	\end{center}
\end{table}

Note that density and the four additional widely-used measures listed in Table~\ref{tab:axioms} do not satisfy all the axioms.
Thus, we propose \textit{overlapness} (see Definition~\ref{defn:overlapness}) as a measure of the degree of hyperedge overlaps, and it satisfies all the axioms,
as formalized in Theorem~\ref{thm:overlapness}.

\begin{defn}[Overlapness] \label{defn:overlapness}
    Given a set of hyperedges $\mathcal{E}$, the overlapness of the set, $o(\mathcal{E})$ is defined as follow:
    \begin{equation*}
        o(\mathcal{E}):=\frac{\sum_{e\in \mathcal{E}}|e|}{|\bigcup_{e \in \mathcal{E}}e|}.
    \end{equation*}
\end{defn}

\begin{thm}[Soundness of Overlapness]\label{thm:overlapness}
    Overlapness $o(\cdot)$ satisfies Axioms~\ref{axm:num_hyperedges}, \ref{axm:num_nodes}, and \ref{axm:size_hyperedges}. 
    
\textsc{\textbf{Proof.}}
See Appendix~\ref{sec:appendix:overlapness_axiom}. \hfill $\blacksquare$
\end{thm}

In overlapness, the sum of sizes of hyperedges, instead of the number of hyperedges, is considered. 
Notably, the overlapness of a hyperedge set is equivalent to the average degree of the distinct nodes in the set. 
In addition, overlapness is equivalent to \textit{weighted} density if we assign the size of each hyperedge as its weight.
Overlapness agrees with our intuition in the previous example.
That is, for $\mathcal{E}_1=\{\{a,b,c\},\{a,b,c,d\},\{a,b,c,d,e\}\}$ and $\mathcal{E}_2=\{\{v,w,x\},\{x,y\},\{y,z\}\}$,
$o(\mathcal{E}_1)=12/5 > o(\mathcal{E}_2)=7/5$. 

\smallsection{Overlapness of Egonets:}
We measure the overlapness of egonets in real-world and randomized hypergraphs, and this leads to Observation~\ref{obs:overlapness}.
As seen in the figures in the second row of Table~\ref{tab:observation:summary}, egonets in real-world hypergraphs tends to have higher overlapness than those in randomized hypergraphs. 
The slopes of the regression lines, which are close to the average egonet overlapness, are steeper in real-world hypergraphs than in randomized ones.

\begin{obs}\label{obs:overlapness}
	Egonets in real-world hypergraphs have higher overlapness than those in randomized hypergraphs. 
\end{obs}

\begin{figure}
	\includegraphics[width=0.18\textwidth]{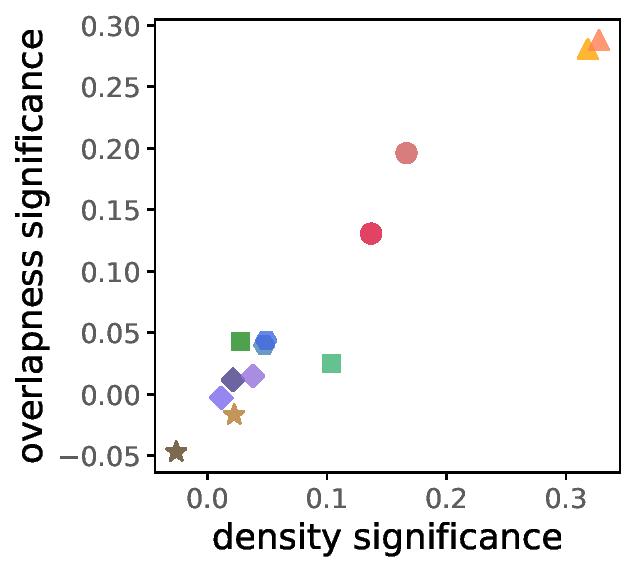}
	\hspace{12pt}
	\includegraphics[width=0.16\textwidth]{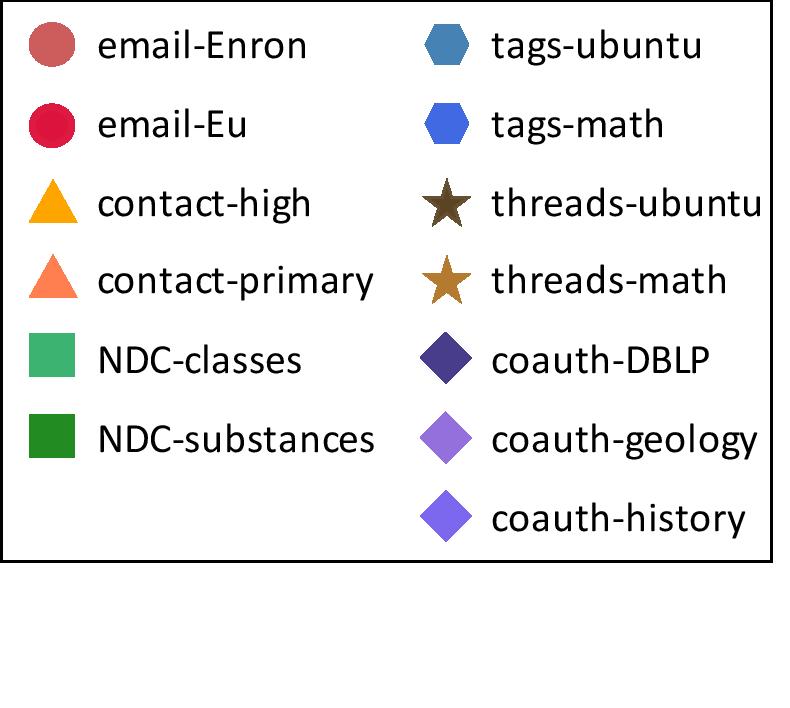}
	\caption{\label{fig:domain}Hypergraphs from the same domain share similar hyperedge overlapping patterns at the egonet level.}
\end{figure}

\smallsection{Comparison across Domains:}
Furthermore, we compute the significance of density and overlapness of egonets in the hypergraph $G$ which are defined as
\begin{align*}
    & \mathrm{sig_{\rho}}(G)\coloneqq\frac{\bar{\rho}(G)-\bar{\rho}(G')}{\max_{\mathrm{g}\in \omega(G),\ \mathrm{g'}\in\omega(G')}|\rho(\mathrm{g})-\rho(\mathrm{g'})|},\\
    & \mathrm{sig_{o}}(G)\coloneqq\frac{\bar{o}(G)-\bar{o}(G')}{\max_{\mathrm{g}\in \omega(G),\ \mathrm{g'}\in\omega(G')}|o(\mathrm{g})-o(\mathrm{g'})|},
\end{align*}
respectively, where $G'$ is a randomized hypergraph of $G$; $\bar{\rho}(\cdot)$ and $\bar{o}(\cdot)$ are the average egonet density and overlapness, respectively; and $\omega(\cdot)$ is the set of egonets. As seen in Figure~\ref{fig:domain}, real-world hypergraphs from the same domain share similar significance of density and overlapness of egonets, indicating that their hyperedges share similar overlapping patterns at the egonet level.

\subsection{L2. Pair/Triple of Nodes Level}\label{sec:observation:pair_triple_level}
Given a pair or triple of nodes, how many hyperedges do overlap at them?
In other words, how many hyperedges do contain the pair or triple?
While the degree is generally defined as the number of hyperedges that contains each individual node, here we extend the concept to pairs and triples of nodes. 
Specifically, if we let $E_S := \{e\in E : S\subseteq e\}$ be the set of hyperedges overlapping at a subset $S\subseteq V$ of nodes, then the degree of each node pair $\{i,j\}$ is defined as $d^{(2)}(\{i,j\}):=|E_{\{i,j\}}|$, and the degree of each node triple $\{i,j,k\}$ is defined as $d^{(3)}(\{i,j,k\}):=|E_{\{i,j,k\}}|$.
The degree of a pair or triple can also be interpreted as the structural similarity between the nodes in the pair or triple. Intuitively, nodes are structurally more similar as they are included together in more hyperedges.

Examining the degree distributions of pairs and triples of nodes, instead of that of individual nodes, gives higher-order insights on how nodes as a set form hyperedges.
In the third and fourth columns of Table~\ref{tab:observation:summary}, we provide the distributions of $d^{(2)}$ and $d^{(3)}$ in real-world hypergraphs and those in a corresponding randomized hypergraph.
Our findings are summarized in Observations~\ref{obs:pair} and \ref{obs:triple}.




\begin{obs}\label{obs:pair}
	The number of hyperedges overlapping at each \textbf{pair of nodes} (i.e., degree of each pair) is more skewed with a heavier tail in real-world hypergraphs than in randomized ones. The distribution is similar to a truncated power law distribution.
\end{obs}

\begin{obs}\label{obs:triple}
	The number of hyperedges overlapping at each \textbf{triple of nodes} (i.e., degree of each triple) is more skewed with a heavier tail in real-world hypergraphs than in randomized ones. The distribution is similar to a truncated power law distribution.
\end{obs}


\begin{table}[t!]
\begin{center}
\caption{\label{loglikelihood-ratio-table} The distribution of the number of hyperedges overlapping at each pair or triple of nodes is heavy-tailed and close to a truncated power-law distribution.
This claim is supported by the reported log-likelihood ratios when fitting the distributions to each of three heavy-tailed distributions (power-law, truncated power-law, and log normal) against the exponential distribution.}
\scalebox{0.86}{
\begin{tabular}{lccccccc}
    \hline
    \multirow{2}{*}{\textbf{Dataset}} & \multicolumn{3}{c}{\textbf{Pair of Nodes (Obs.~\ref{obs:pair})}} && \multicolumn{3}{c}{\textbf{Triple of Nodes (Obs.~\ref{obs:triple})}}\\ \cline{2-4}\cline{6-8}
    & pw & tpw & logn & & pw & tpw & logn \\\hline
    email-Enron & -0.36 & \textbf{\underline{\smash{4.22}}} & \textbf{3.50} & & \textbf{1.91} & \textbf{\underline{\smash{3.88}}} & \textbf{3.47}\\
	email-Eu & \textbf{0.66} & \textbf{\underline{\smash{1.48}}}  & \textbf{1.29} & & \textbf{0.21} & \textbf{\underline{\smash{0.77}}} & \textbf{0.63}\\
	contact-primary & \textbf{0.64} & \textbf{\underline{\smash{1.40}}}  & \textbf{1.35} & & \textbf{0.01} & \textbf{\underline{\smash{0.48}}} & \textbf{\underline{\smash{0.48}}}\\
	contact-high & \textbf{0.75} & \textbf{\underline{\smash{0.81}}} & \textbf{0.79} & & -1.04 & - & \textbf{\underline{\smash{0.80}}} \\
	NDC-classes & \textbf{13.49} & \textbf{\underline{\smash{15.74}}} & \textbf{14.78} & & \textbf{24.37} & \textbf{\underline{\smash{31.53}}} & \textbf{29.19} \\
	NDC-substances & \textbf{38.68} & \textbf{\underline{\smash{43.87}}}  & \textbf{42.55}  & & \textbf{102.90} & \textbf{\underline{\smash{116.45}}} & \textbf{109.77} \\
	tags-ubuntu & \textbf{39.66} & \textbf{\underline{\smash{41.55}}} & \textbf{41.25} & & \textbf{17.03} & \textbf{\underline{\smash{17.84}}} & \textbf{17.79} \\
	tags-math & \textbf{3.82} & \textbf{\underline{\smash{4.49}}}  & \textbf{4.47} & & \textbf{26.97} & \textbf{\underline{\smash{29.26}}} & \textbf{29.07} \\
	threads-ubuntu & \textbf{3.79} & \textbf{\underline{\smash{3.97}}}  & \textbf{3.97} & & \textbf{0.34} & \textbf{\underline{\smash{0.80}}} & \textbf{0.73}\\
	threads-math & \textbf{14.25} & \textbf{\underline{\smash{14.78}}} & \textbf{14.68} & & -1.04 & -0.09 & -1.12 \\
 	coauth-DBLP & \textbf{19.23} & \textbf{\underline{\smash{22.47}}} & \textbf{22.31} & & \textbf{5.75} & \textbf{\underline{\smash{5.84}}} & \textbf{5.83}\\
 	coauth-geology & \textbf{45.20} & \textbf{\underline{\smash{53.39}}} & \textbf{52.92} & & \textbf{9.69} & \textbf{\underline{\smash{13.73}}} & \textbf{13.01} \\
 	coauth-history & \textbf{3.74} & \textbf{3.81} & \textbf{\underline{\smash{3.91}}} & & -0.36 & \textbf{\underline{\smash{1.42}}} & \textbf{1.27}\\
    \hline
\end{tabular}}
\end{center}
\end{table}

In addition to the visual inspection, we compute the log-likelihood ratio of three representative heavy-tailed distributions (power-law, truncated power-law, and log normal) against the exponential distribution, as suggested in \cite{alstott2014powerlaw,clauset2009power}.
If the ratio is greater than $0$, the given distribution is more similar to the corresponding heavy-tailed distribution than an exponential distribution.
As reported in Table~\ref{loglikelihood-ratio-table}, except for one case, at least one heavy-tailed distribution has a positive ratio, and in most cases the ratio is highest for truncated power-law distributions.
These result support the claim that the degree distributions of pairs and triples of nodes is heavy-tailed and similar to truncated power-law distributions.

In fact, these results are intuitive. The more often a pair or triple of nodes interact together, the more likely they are to interact together again. 
For example, researchers that have co-authored multiple papers are likely to share common interests, which can lead to more collaborations in the future.

\begin{table}[t!]
\begin{center}
\caption{\label{loglikelihood-ratio-table-homogeneity} The distributions of hyperedge homogeneity in  real hypergrpahs and those generated by \methodauto are heavy-tailed. Log-likelihood ratios are calculated as in Table~\ref{loglikelihood-ratio-table}. }
\scalebox{0.86}{
\begin{tabular}{lccccccc}
    \hline
    \multirow{2}{*}{\textbf{Dataset}} & \multicolumn{3}{c}{Real-World Data} & & \multicolumn{3}{c}{Generated} \\ \cline{2-4}\cline{6-8}
    & pw & tpw & logn & & pw & tpw & logn \\\hline
    email-Enron & -1.09 & -0.26 & -0.38 & & -2.71 & -0.43 & -4.76\\
	email-Eu & \textbf{0.90} & \textbf{0.90}  & \textbf{\underline{\smash{0.91}}} & & -3.00 & \textbf{\underline{\smash{3.13}}} & \textbf{2.08} \\
	contact-primary & \textbf{2.19} & \textbf{\underline{\smash{2.30}}}  & \textbf{2.22} & & \textbf{0.67} & \textbf{\underline{\smash{2.26}}} & \textbf{1.90}\\
	contact-high & \textbf{1.55} & \textbf{1.55} & \textbf{\underline{\smash{1.95}}} & & \textbf{2.50} & \textbf{\underline{\smash{4.72}}} & \textbf{3.65}\\
	NDC-classes & 0.00 & \textbf{\underline{\smash{0.39}}} & \textbf{0.18} & & -0.47 & \textbf{\underline{\smash{0.87}}} & \textbf{0.52}\\
	NDC-substances & \textbf{0.64} & \textbf{\underline{\smash{1.22}}}  & \textbf{1.13} & & \textbf{1.87} & \textbf{\underline{\smash{2.90}}} & \textbf{2.58}\\
	tags-ubuntu & \textbf{2.25} & \textbf{2.25} &  \textbf{\underline{\smash{2.26}}} & & -2.01 & \textbf{\underline{\smash{7.00}}} & \textbf{6.19}\\
	tags-math & -17.66 & -7.93 & \textbf{\underline{\smash{2.62}}} & & \textbf{3.53} & \textbf{\underline{\smash{6.56}}} & \textbf{6.07} \\
	threads-ubuntu & \textbf{4.58} & \textbf{\underline{\smash{7.70}}}  & \textbf{6.55} & & \textbf{3.92} & \textbf{\underline{\smash{4.25}}} & \textbf{3.94}\\
	threads-math & -0.72 & \textbf{\underline{\smash{9.00}}} & \textbf{6.69} & & \textbf{4.30} & \textbf{\underline{\smash{12.10}}} & \textbf{10.53} \\
 	coauth-DBLP & \textbf{4.01} & \textbf{\underline{\smash{4.31}}} & \textbf{4.20} & & \textbf{10.65} & \textbf{\underline{\smash{25.23}}} & \textbf{22.82}\\
 	coauth-geology & \textbf{4.29} & \textbf{\underline{\smash{5.52}}} & \textbf{5.37} & & \textbf{1.75} & \textbf{\underline{\smash{8.06}}} & \textbf{7.00}\\
 	coauth-history & - & - & \textbf{\underline{\smash{1.73}}} & & \textbf{3.98} & \textbf{\underline{\smash{4.31}}} & \textbf{4.02}\\
    \hline
\end{tabular}}
\end{center}
\end{table}

\subsection{L3. Hyperedge Level}~\label{sec:observation:hyperedge_level}

How are nodes that form hyperedges together related to each other?
It is unlikely in real-world hypergraphs that each hyperedge is formed by nodes chosen independently at random.
It is expected to exist a strong dependency among the nodes forming a hyperedge together. 
In order to investigate the dependency, we use the \textit{homogeneity} of hyperedge, defined in Definition~\ref{defn:homogeneity}, to measure how structurally similar such nodes are.

\begin{defn}[Homogeneity of a Hyperedge] \label{defn:homogeneity}
    The homogeneity of a hyperedge $e\in E$ is defined as follow:
    \begin{equation}
        homogeneity(e):=
            \begin{cases}
                \frac{\sum_{\{u,v\}\in \binom{e}{2}}|E_{\{u,v\}}|}{\binom{|e|}{2}},& \mathrm{if}\ |e| > 1\\
                0,& \mathrm{otherwise,}
            \end{cases} \label{eq:homo}
    \end{equation}
\end{defn}
\noindent where $\binom{e}{2}$ is the set of node pairs in $e$ and $|E_{\{u,v\}}|$ is the number of hyperedges overlapping at the pair of $u$ and $v$ (i.e., the degree of the pair $\{u,v\}$).
Note that, in Eq.~\eqref{eq:homo}, the structural similarity between two nodes is measured in terms of the number of hyperedges overlapping at them, which we examine in Section~\ref{sec:observation:pair_triple_level}.
Eq.~\eqref{eq:homo} can be easily extended to three or more nodes. 

The figures in the last row of Table~\ref{tab:observation:summary} show the homogeneity of the hyperedges in real-world hypergraphs and corresponding randomized hypergraphs. As summarized in Observation~\ref{obs:hyperedge_locality}, there is a tendency that the homogeneity of each hyperedge in real-world hypergraphs is greater than that in randomized ones.
Moreover, we verify that the distribution of homogeneity is heavy-tailed (see Table~\ref{loglikelihood-ratio-table-homogeneity}), as in the previous subsection.

\begin{obs}\label{obs:hyperedge_locality}
	Hyperedges in real-world hypergraphs tend to contain structurally more similar nodes (i.e., nodes where many hyperedges overlap) than those in randomized hypergraphs. 
\end{obs}

The homogeneity of hyperedges plays a key role in generating realistic hypergraphs, as described in the following section.

%% file: 050method.tex
We have shown that overlapping patterns of hyperedges in real-world hypergraphs are clearly distinguished from those in randomized hypergraphs. In this section, we propose \method, a scalable and realistic hypergraph generative model that reproduces the realistic overlapping patterns of hyperedges. 
After describing \method, we present \methodauto, which automatically tunes the parameters of \method so that hypergraphs similar to a given target hypergraph are generated.
Then, we evaluate \method and \methodauto experimentally.


\subsection{\method: Multilevel \hypercl~\label{sec:method:our_method}}
We propose \method, a realistic hypergraph generative model whose pseudocode is described in Algorithm~\ref{alg:method}. The key idea behind \method is to
extend \hypercl to multiple levels.
Recall that \hypercl itself cannot accurately reproduce realistic overlapping patterns, as shown in Section~\ref{sec:observation}.

\begin{algorithm}[t]
	\caption{\method: Realistic Hypergraph Generator \label{alg:method}}
	\SetKwInOut{Input}{Input}
    \SetKwInOut{Output}{Output}
    \Input{ (1) distribution of hyperedge sizes $\{s_1,\cdots,s_{|E|}\}$\\ (2) distribution of node degrees $\{d_1,\cdots,d_{|V|}\}$\\ (3) number of levels $L$ ($\leq \log_{2}|V|$)\\ (4) weights of each level $\{w_1,\cdots,w_{L}\}$}
    \Output{ synthetic hypergraph $\hat{G}=(\hat{V},\hat{E})$}
    \vspace{3pt}
    
    \texttt{\color{blue}/* Initialization */}\\
    $\hat{V}\leftarrow \{1,\cdots,|V|\}$ and
    $\hat{E}\leftarrow \varnothing $ \\
    \vspace{3pt}
    
    \texttt{\color{blue}/* Hierarchical Node Partitioning */}~\label{alg:method:node_partition}\\
    $S_{1}^{(L)},\cdots,S_{2^{L-1}}^{(L)}\leftarrow$ uniformly partition $\hat{V}$ into $2^{L-1}$ groups\\
    \For{\upshape\textbf{each} level $\ell=L-1,\cdots,1$}{
        \For{\upshape\textbf{each} group $\textsl{g}=1,\cdots,2^{\ell-1}$}{
            $S_{g}^{(\ell)}=S_{2g-1}^{(\ell+1)}\cup S_{2g}^{(\ell+1)}$ \label{alg:method:node_partition:union}\\
        }
    }
    \vspace{3pt}
    
    \texttt{\color{blue}/* Hyperedge Generation */} \label{alg:method:hyperedge_generation}\\
    \For{\upshape\textbf{each} $i=1,\cdots,|E|$}{
        $\ell \leftarrow$ select a level with prob. proportional to the weight \\
        $S_{g}^{(\ell)} \leftarrow$ select a group at level $\ell$ uniformly at random \\
        $\hat{e_i}\leftarrow \varnothing$\\
        \While{$|\hat{e_i}|<s_i$}{
            $v\leftarrow$ select a node from $S_{g}^{(\ell)}$ with prob. proportional to the degree 
            \label{alg:method:hyperedge_generation:select:node}
            \\
            $\hat{e_i}=\hat{e_i}\cup \{v\}$\\
        }
        $\hat{E}=\hat{E}\cup \{\hat{e_i}\}$\label{alg:method:hyperedge_generation:include_hyperedge}\\
    }
    \vspace{3pt}
    
    \Return{$\hat{G}=(\hat{V},\hat{E})$}
\end{algorithm}

\smallsection{Description of \method:}
\method, a multilevel extension of \hypercl, requires two additional inputs: (1) number of levels $L$ and (2) weights of each level $\{w_1,\cdots,w_L\}$.\footnote{$L$ should be set such that $L\leq\log_{2}{|V|}$.} For now, we assume that the parameters are given; how to set the parameters is discussed in the next subsection. \method consists of the hierarchical node partitioning step and the hyperedge generation step.

\vspace{5pt}
\noindent\textbf{Step 1. Hierarchical Node Partitioning (lines~\ref{alg:method:node_partition} -~\ref{alg:method:node_partition:union}).}
\method first partitions nodes into groups at every level.
Specifically, at every level $\ell \in \{1,\cdots,L\}$, it randomly divides nodes $2^{\ell-1}$ groups, denoted by $S_{1}^{(\ell)}, \cdots, S_{2^{\ell-1}}^{(\ell)}$ while satisfying the following conditions:
\setlength{\leftmargini}{1.75em}
\begin{enumerate}
    \item $S_{i}^{(\ell)}\cap S_{j}^{(\ell)}=\varnothing$ for all $i\neq j\in \{1,\cdots ,2^{\ell-1}\}$,
    \item $|\bigcup_{i=1}^{2^{\ell-1}}S_{i}^{(\ell)}|=|V|$, 
     \item $|S_{i}^{(\ell)}|=\lfloor\frac{|V|\cdot i}{2^{\ell-1}}\rfloor -\lfloor\frac{|V|\cdot (i-1)}{2^{\ell-1}}\rfloor$ for all $i\in \{1,\cdots ,2^{\ell-1}\}$, 
    \item $S_{i}^{(\ell)}=S_{2i-1}^{(\ell+1)}\cup S_{2i}^{(\ell+1)}$ for all $\ell<L$, and $i\in \{1,\cdots, 2^{\ell-1}\}$.
\end{enumerate}
The first and second conditions ensure that at each level each node belongs to exactly one group.
The third condition states that the size of groups at each level are almost uniform.
The last condition states that the groups are hierarchical. That is, if nodes are in the same group at a level, then they are in the same group at all lower levels.
Note that nodes are divided more finely into smaller subsets at higher levels. At the lowest level $1$, there exist a single group, which is the same as the entire set of nodes $V$, whereas at the highest level $L$, there exist most groups whose number is $2^{L-1}$.

\vspace{5pt}
\noindent\textbf{Step 2. Hyperedge Generation (lines~\ref{alg:method:hyperedge_generation} -~\ref{alg:method:hyperedge_generation:include_hyperedge}).}
Once we partition nodes hierarchically in the previous step, for each $i$-th hyperedge $\hat{e_i}$, \method first selects a level with probability proportional to the weight of each level. 
That is, each level $\ell$ is selected with probability proportional to $w_\ell$.
At the selected level $\ell$, \method selects a group $S_{g}^{(\ell)}$ uniformly at random. 
Then, the nodes forming $\hat{e_i}$ are sampled independently, with probability proportional to the degree of each node,\footnote{For each node $v\in S_{g}^{(\ell)}$, the probability is $d_v/\sum_{j\in S_{g}^{(\ell)}}d_j$.} until the size of the hyperedge reaches $s_i$.
That is, instead of taking all nodes into consideration, we divide the nodes into multiple groups and limit the nodes that a hyperedge can contain into those in a group.
Note that hyperedges generated within the same group at higher levels are more likely to be overlapped each other, as fewer nodes are in each group at a higher level.
\blue{Practically, since $\hat{e_i}$ cannot be generated from a group whose size is smaller than $s_i$, we select level $\ell$ such that $\ell \leq \log_2{\frac{|V|}{s_i}}+1$.}

\smallsection{Degree Preservation of \method:}
In hypergraphs generated by \method, the size distribution of hyperedges is exactly the same as the input size distribution.
Specifically, $|\hat{e_i}|=s_i$ holds for all $i\in \{1,\cdots,|E|\}$.
The degree distribution of nodes is also expected to be similar to the input degree distribution.
In order to show this, we first provide Lemma~\ref{lma:group_selection}, which our analysis is based on.
\begin{lma}~\label{lma:group_selection}
     For each group $S_{g}^{(\ell)}$ at level $\ell$, the probability for a hyperedge $e$ to be generated from $S_{g}^{(\ell)}$ is
    \begin{equation}~\label{eq:group_selection}
       P[e\subseteq S_{g}^{(\ell)}]= \frac{w_{\ell}}{W_e}\cdot \frac{1}{2^{\ell-1}},
    \end{equation}
    where $W_e$ is \blue{the sum of the weights of suitable levels.} That is, $W_e=\sum_{k=1}^{L_e}w_k$ where $L_e=\lfloor\log_2{\frac{|V|}{s_i}+1}\rfloor$.
\end{lma}
\textsc{\textbf{Proof.}} 
Given any hyperedge $e$, \method first randomly selects a \blue{suitable} level with probability proportional to the given weight. 
Thus, the probability for the level $\ell$ to be selected is $w_{\ell}/W_e$. 
Once the level is determined, any of the $2^{\ell-1}$ groups in level $\ell$ is selected uniformly at random, i.e., with probability $1/2^{\ell-1}$.
The probability for $e$ to be generated from $S_{g}^{(\ell)}$ is the product of the two probabilities, and thus Eq.~\eqref{eq:group_selection} holds. \hfill $\blacksquare$
    

For each node $v$, let $\hat{d}_v^{(\ell)}$ be the number of hyperedges that contain the node $v$ among those generated at level $\ell$.
Then, the degree $\hat{d}_v$ of $v$ in an output hypergraph is the sum of $\hat{d}_v^{(\ell)}$ over all levels, i.e., $\hat{d}_v=\sum_{\ell=1}^{L}\hat{d}_v^{(\ell)}$. 
Let $d_{max}:=\max_{k\in  \{1,\cdots,|V|\}}d_{k}$ and $s_{max}=\max_{k\in  \{1,\cdots,|E|\}}s_{k}$.
Assume  $|V| \gg 2^{L-1} \cdot d_{max}$ and $\sum_{j\in S_{g}^{(\ell)}} d_j \gg  d_{max} \cdot s_{max}$ for all $S_{g}^{(\ell)}$.\footnote{If $|V| \gg 2^{L-1} \cdot d_{max}$, then $\sum_{k\in S_{g}^{(\ell)}} d_k/\sum_{j=1}^{|V|}d_j\approx 1/2^{\ell-1}$ for all $\ell\in\{1,\cdots,L\}$.}
Then,
\vspace{-3mm}
\begin{align*}
        \mathbb{E}[\hat{d}_v] &= \sum_{\ell=1}^{L}\mathbb{E}[\hat{d}_v^{(\ell)}] = \sum_{\ell=1}^{L} \sum_{e\in E} P[e \subseteq S_{g}^{(\ell)}(v)] \cdot P[v\in e |e \subseteq S_{g}^{(\ell)}(v)] \\
	    &\approx\sum_{e\in E}\sum_{\ell=1}^{L_e}\left(\frac{w_{\ell}}{W_e} \cdot \frac{1}{2^{\ell-1}} \right) \left[|e| \cdot \left(\frac{d_v \cdot 2^{\ell-1}}{\sum_{j=1}^{|V|}d_j}\right)\right]\\
		& =\frac{d_v}{\sum_{j=1}^{|V|}d_j} \cdot \sum_{e\in E}\left(|e|\cdot \sum_{\ell=1}^{L_e}\frac{w_{\ell}}{W_e}\right)=
		\blue{d_v\cdot \frac{\sum_{e\in E}|e|}{\sum_{j=1}^{|V|}d_j}} = 
		d_v, 
\end{align*}
where $S_{g}^{(\ell)}(v)$ is the group at level $\ell$ containing $v$.
That is, $\hat{d}_v$ is expected to be close to $d_v$, as we confirm empirically in \cite{online2020appendix}. 


\begin{algorithm}[t]
    \DontPrintSemicolon
	\caption{\methodauto: Automatic Parameter Selection \label{alg:optimization}}
	\SetKwInOut{Input}{Input}
    \SetKwInOut{Output}{Output}
    \SetKwFunction{algo}{algo}\SetKwFunction{proc}{update}
    
    \Input{(1) input hypergraph $G=(V,E)$\\(2) update resolution $p$}
    \Output{synthetic hypergraph $\hat{G}=(\hat{V},\hat{E})$}
    $\hat{G}=(\hat{V},\hat{E})\leftarrow$ run \hypercl using the distributions in $G$ \label{alg:auto:init}\\
    \For{\upshape\textbf{each} level $\ell=2,\cdots,L$}{
        $i^{*}\leftarrow \argmin_{i\in\{1,\cdots,1/p\}}HHD\left(G, \mathtt{update}(\hat{G},p\cdot i, \ell)\right)$ \label{alg:auto:opt}\\
        $\bar{G}\leftarrow \mathtt{update}(\hat{G}, p\cdot i^{*}, \ell)$\\
        \lIf{$HHD(G, \bar{G}) < HHD(G, \hat{G})$}{
            $\hat{G}\leftarrow \bar{G}$\label{alg:auto:update}
        } \lElse{
            \textbf{break}\label{alg:auto:break}
        }
    }
    \Return{$\hat{G}=(\hat{V},\hat{E})$}
    
    \vspace{1pt}
    \setcounter{AlgoLine}{0}
    \SetKwProg{subproc}{}{}{}
    \subproc{\proc{$\hat{G}=(\hat{V},\hat{E})$, $q$, $\ell$}}{
        $\bar{G}(\bar{V},\bar{E})\leftarrow \hat{G}(\hat{V},\hat{E})$\\
        remove $(q\cdot 100)\%$ of the hyperedges created at level $\ell-1$ \\
        create the same number of hyperedges at level $\ell$\\
        \Return{$\bar{G}=(\bar{V},\bar{E})$}
    }
\end{algorithm}

\begin{table*}[t!]
\vspace{-3mm}
\begin{center}
\caption{\label{tab:density_overlapness_homogeneity} D-statistics between the distributions of (1) egonet density, (2) egonet overlapness and (3) hyperedge homogeneity in real-world hypergraphs and corresponding hypergraphs generated by five models: \hypercl (\textsc{H-CL}), \hyperpa (\textsc{H-PA}), \hyperff (\textsc{H-FF}), \method (\textsc{H-LAP}), and \methodauto (\textsc{H-LAP}\textsuperscript{+}). \methodauto reproduces the distributions most accurately.
}
\scalebox{0.885}{
\begin{tabular}{lccccccccccccccccc}
    \hline
    \multirow{2}{*}{\textbf{Dataset}} & \multicolumn{5}{c}{\textbf{Density of Egonets (Obs.~\ref{obs:density})}} && \multicolumn{5}{c}{\textbf{Overlapness of Egonets (Obs.~\ref{obs:overlapness})}} &&  \multicolumn{5}{c}{\textbf{Homogeneity of Hyperedges (Obs.~\ref{obs:hyperedge_locality})}} \\\cline{2-6}\cline{8-12}\cline{14-18}
    & \small{\textsc{H-CL}} & \small{\textsc{H-PA}} & \small{\textsc{H-FF}} & \small{\textsc{H-LAP}} & \small{\textbf{\textsc{H-LAP}\textsuperscript{+}}} & & \small{\textsc{H-CL}} & \small{\textsc{H-PA}} & \small{\textsc{H-FF}} & \small{\textsc{H-LAP}} & \small{\textbf{\textsc{H-LAP}\textsuperscript{+}}} & & \small{\textsc{H-CL}} & \small{\textsc{H-PA}} & \small{\textsc{H-FF}} & \small{\textsc{H-LAP}} & \small{\textbf{\textsc{H-LAP}\textsuperscript{+}}}\\\hline
	email-Enron &  0.545 & 0.202 & 0.391 & 0.405 & \textbf{0.125} & & 0.517 & 0.398 & 0.398 & 0.391 & \textbf{0.111} & & 0.498 & 0.241 & 0.656 & 0.191 & \textbf{0.136}\\
    email-Eu & 0.724 & - & 0.402 & 0.577 & \textbf{0.310} & & 0.534 & - & 0.639 & 0.432 & \textbf{0.197} & & 0.505 & - & 0.688 & 0.247 & \textbf{0.168}\\
    contact-primary & 0.896 & 0.537 & 0.975 & 0.334 & \textbf{0.128} & & 0.867 & 0.471 & 0.942 & 0.285 & \textbf{0.095} & & 0.430 & 0.236 & 0.484 & \textbf{0.142} & 0.188\\
    contact-high & 0.948 & 0.529 & 0.880 & 0.522 & \textbf{0.345} & & 0.874 & 0.431 & 0.703 & 0.486 & \textbf{0.296} & & 0.423 & 0.196 & 0.336 & \textbf{0.120} & 0.178\\
    NDC-classes & 0.694 & 0.785 & 0.731 & 0.696 & \textbf{0.635} & & 0.302 & 0.715 & 0.406 & \textbf{0.231} & 0.248 & & 0.274 & 0.410 & 0.484 & 0.272 & \textbf{0.225}\\
    NDC-substances & 0.451 & - & 0.801 & 0.426 & \textbf{0.366} & & 0.321 & - & 0.338 & 0.243 & \textbf{0.157} & & 0.377 & - & 0.740 & 0.262 & \textbf{0.108}\\
    tags-ubuntu & 0.522 & \textbf{0.162} & 0.216 & 0.410 & 0.300 & & 0.432 & \textbf{0.117} & 0.398 & 0.487 & 0.210 & & 0.245 & 0.136 & 0.844 & 0.105 & \textbf{0.011}\\
    tags-math & 0.496 & 0.350 & 0.561 & \textbf{0.195} & 0.227 & & 0.460 & 0.325 & 0.709 & \textbf{0.151} & 0.186 & & 0.337 & 0.217 & 0.921 & 0.086 & \textbf{0.015}\\
    threads-ubuntu & 0.159 & 0.856 & - & 0.163 & \textbf{0.159} & & 0.299 & 0.953 & - & 0.300 & \textbf{0.297} & & 0.020 & 0.291 & - & 0.016 & \textbf{0.011}\\
    threads-math & 0.137 & 0.492 & - & \textbf{0.120} & 0.135 & & 0.232 & 0.714 & - & 0.235 & \textbf{0.229} & & 0.060 & 0.368 & - & 0.102 & \textbf{0.019}\\
    coauth-DBLP & 0.228 & - & - & 0.227 & \textbf{0.132} & & 0.302 & - & - & 0.267 & \textbf{0.244} & & 0.715 & - & - & 0.540 & \textbf{0.026}\\
    coauth-geology & 0.200 & - & - & 0.202 & \textbf{0.138} & & \textbf{0.248} & - & - & 0.252 & 0.266 & & 0.624 & - & - & 0.481 & \textbf{0.044}\\
    coauth-history & \textbf{0.087} & - & - & 0.090 & 0.089 & & \textbf{0.316} & - & - & 0.321 & 0.324 & & 0.154 & - & - & 0.125 & \textbf{0.020}\\
    \hline
    \textbf{Average} & 0.468 & 0.489 & 0.619 & 0.335 & \textbf{0.237} & & 0.439 & 0.515 & 0.566 & 0.313 & \textbf{0.219} & & 0.358 & 0.261 & 0.644 & 0.206 & \textbf{0.088}\\
    \hline
    \multicolumn{7}{l}{-: out of time (taking more than 10 hours) or out of memory} \\
\end{tabular}}
\end{center}
\end{table*}

\begin{table*}[t!]
\vspace{-4mm}
\begin{center}
\caption{\label{tab:experiment:pair} Distributions of the number of overlapping hyperedges at each pair and each triple of nodes are reproduced accurately by \textcolor{myblue}{\methodauto}, while \textcolor{myred}{\hypercl} fails in many cases. They obey heavy-tailed distribution, as in the \textcolor{mygreen}{real ones}.}
\scalebox{0.95}{
\begin{tabular}{c|cccccc}
\toprule
 & \textbf{email-Eu} & \textbf{contact-primary} & \textbf{NDC-substances} & \textbf{tags-math} & \textbf{threads-ubuntu} & \textbf{coauth-DBLP}\\ 
 \hline 
\parbox[t]{2mm}{\multirow{6}{*}{\rotatebox[origin=c]{90}{\ \ Observation~\ref{obs:pair}}}} & \raisebox{-\totalheight}{\includegraphics[width=0.1475\textwidth]{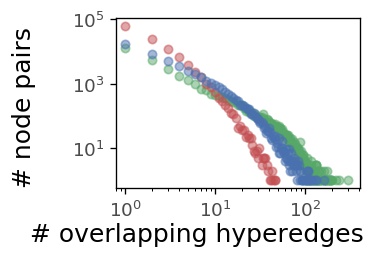}} & \raisebox{-\totalheight}{\includegraphics[width=0.1475\textwidth,]{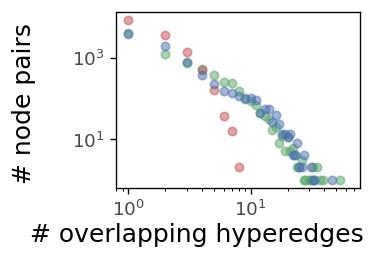}} & \raisebox{-\totalheight}{\includegraphics[width=0.1475\textwidth]{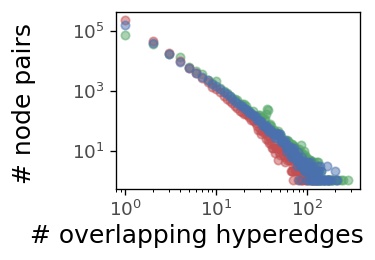}} & \raisebox{-\totalheight}{\includegraphics[width=0.1475\textwidth]{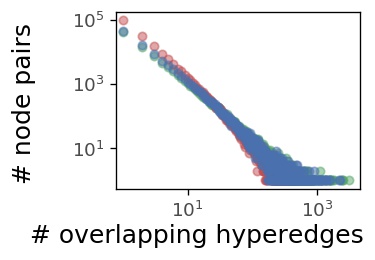}} & \raisebox{-\totalheight}{\includegraphics[width=0.1475\textwidth]{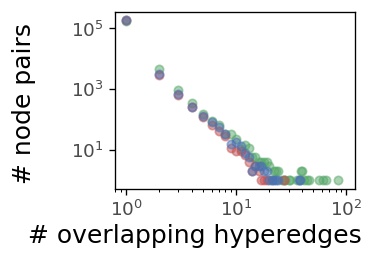}} & \raisebox{-\totalheight}{\includegraphics[width=0.1475\textwidth]{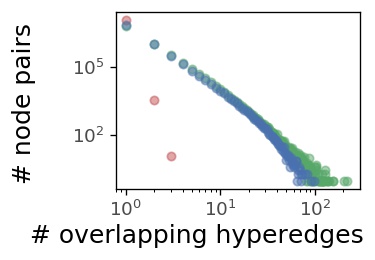}}\\
 \hline
\parbox[t]{2mm}{\multirow{6}{*}{\rotatebox[origin=c]{90}{\ \ Observation~\ref{obs:triple}}}} & \raisebox{-\totalheight}{\includegraphics[width=0.1475\textwidth]{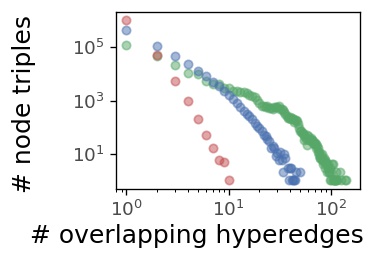}} & \raisebox{-\totalheight}{\includegraphics[width=0.1475\textwidth,]{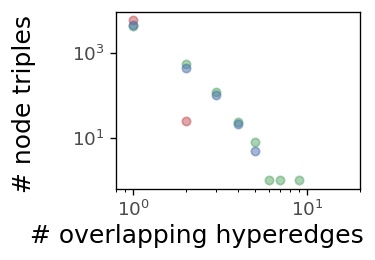}} & \raisebox{-\totalheight}{\includegraphics[width=0.1475\textwidth]{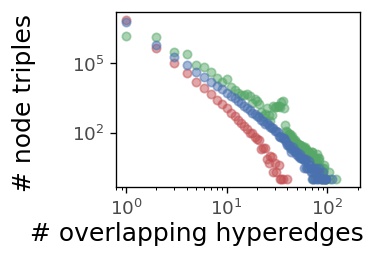}} & \raisebox{-\totalheight}{\includegraphics[width=0.1475\textwidth]{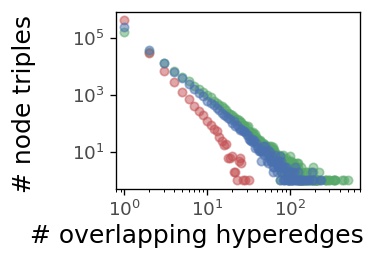}} & \raisebox{-\totalheight}{\includegraphics[width=0.1475\textwidth]{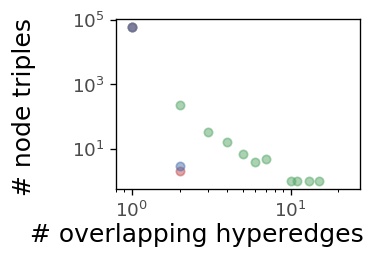}} & \raisebox{-\totalheight}{\includegraphics[width=0.1475\textwidth]{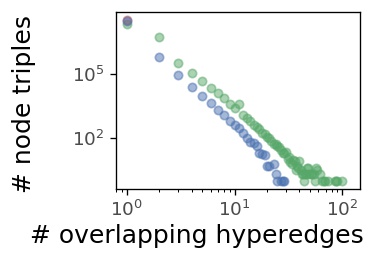}}\\
\bottomrule
 \end{tabular}}
\end{center}
\end{table*}

\smallsection{Intuition Behind \method:}
In this section, we provide some reasons why we expect \method to accurately reproduce the realistic overlapping patterns of hyperedges discovered in Section~\ref{sec:observation}.

\setlength{\leftmargini}{1em}
\begin{itemize}
    \item For a pair or triple of nodes belonging to the same small group, the number of hyperedges overlapping at them is expected to be high. Thus, the distribution of the number of overlapping hyperedges at each pair or triple is expected to be skewed. 
	\item As hyperedges can be formed within a small group, which contains structurally similar nodes, the homogeneity of each hyperedge is expected to be high. Moreover, as the size of groups varies, the homogeneity of hyperedges is expected to vary depending on the size of the groups that they are generated from.
	\item As the hyperedges in the egonet of each node $v$ are likely to contain nodes belonging to the same small group with $v$, their density and overlapness are expected to be high. \ \\
\end{itemize}


\subsection{\methodauto: Parameter Selection}

Given an input hypergraph $G$, how can we set the parameters of \method (i.e., the number of levels $L$ and the weight of each level $\{w_1,\cdots,w_L\}$) so that it generates a synthetic hypergraph $\hat{G}$ especially similar to a target real-world hypergraph? The parameters should be carefully tuned since the structural properties of the generated hypergraphs vary depending on their settings. %
To this end, we propose \methodauto, which automatically tunes the parameters.

\smallsection{Hyperedge Homogeneity Objective:} 
As its objective function, \methodauto uses the \textit{hyperedge homogeneity distance} $HHD(G,\hat{G})$ between the input hypergraph $G$ and a generated hypergraph $\hat{G}$. It is defined as the \blue{Kolmogorov-Smirnov D-statistics} between the hyperedge homogeneity distribution of $G$ and that of $\hat{G}$. That is, 
\begin{equation}
    HHD(G,\hat{G}) = \max_x \{|F(x)-F'(x)|\}, \label{eq:objective}
\end{equation}
where $F$ and $F'$ are the cumulative hyperedge homogeneity distribution of hypergraph $G$ and $\hat{G}$, respectively. Then, assuming that the number of levels $L$ is given, \methodauto aims to find the weights of levels that minimize the hyperedge homogeneity distance. That is, \methodauto aims to solve the following optimization problem:
\begin{equation*}
    \min_{w_1,\cdots,w_L} HHD(G, \hat{G}),
\end{equation*}
where we assume $w_1+\cdots+w_L=1$ since only the ratios between the weights matter.

\begin{table*}[t!]
	\vspace{-4mm}
	\begin{center}
		\caption{\label{tab:pair_triple}
			D-statistics between the distributions of the number of overlapping hyperedges at each pair and each triple of nodes in real-world hypergraphs and corresponding hypergraphs generated by five models: \hypercl (\textsc{H-CL}), \hyperpa (\textsc{H-PA}), \hyperff (\textsc{H-FF}), \method (\textsc{H-LAP}), and \methodauto (\textsc{H-LAP}\textsuperscript{+}). \methodauto reproduces the distributions most accurately, \blue{and these distributions follow heavy-tailed distributions.}}
		\scalebox{0.835}{
			\begin{tabular}{lccccccccccccccccccc}
				\hline
				\multirow{3}{*}{\textbf{Dataset}} & \multicolumn{9}{c}{\textbf{Pair of Nodes (Obs.~\ref{obs:pair})}} && \multicolumn{9}{c}{\textbf{Triple of Nodes (Obs.~\ref{obs:triple})}} \\\cline{2-10}\cline{12-20}
				& \multicolumn{5}{c}{Distance from Real (D-statistics)} & &  \multicolumn{3}{c}{Heavy-tail Test} & & \multicolumn{5}{c}{Distance from Real (D-statistics)} &  &\multicolumn{3}{c}{Heavy-tail Test}\\\cline{2-6}\cline{8-10}\cline{12-16}\cline{18-20}
				& \small{\textsc{H-CL}} & \small{\textsc{H-PA}} & \small{\textsc{H-FF}} & \small{\textsc{H-LAP}} & \small{\textbf{\textsc{H-LAP}\textsuperscript{+}}} & & pw & tpw & logn & & \small{\textsc{H-CL}} & \small{\textsc{H-PA}} & \small{\textsc{H-FF}} & \small{\textsc{H-LAP}} & \small{\textbf{\textsc{H-LAP}\textsuperscript{+}}} & & pw & twp & logn \\\hline
				email-Enron & 0.143  & \textbf{0.056}  & 0.217 & 0.075 & 0.139  & & -2.37 & -0.29 & -1.53 & & 0.089  & 0.295 & 0.136 & \textbf{0.061} & 0.072 & & -0.22 & \textbf{0.38} & \textbf{0.24}\\
				email-Eu & 0.225 & - & 0.352 & 0.162 & \textbf{0.066} & & \textbf{0.24} & \textbf{2.75} & \textbf{2.53} & &  0.480  & - & 0.516 & 0.337 & \textbf{0.206} & & \textbf{0.41} & \textbf{2.11} & \textbf{1.96}\\
				contact-primary & 0.196  & 0.062  & 0.223 & 0.070 & \textbf{0.051} & & \textbf{9.53} & \textbf{15.74} & \textbf{13.92} & & 0.137 & 0.061  & 0.110 & 0.053  & \textbf{0.031} & & -1.86 & -1.27 & \textbf{1.23}\\
				contact-high & 0.277 & \textbf{0.062} & 0.141  & 0.127 & 0.067 & & -3.09 & -0.95 & -0.06 & & 0.210 & \textbf{0.131} & 0.182 & 0.182 & 0.193 & & -3.95 & - & \textbf{0.50}\\
				NDC-classes & 0.273 & 0.197 & 0.196 & 0.246 & \textbf{0.172} & & \textbf{12.15} & \textbf{14.42} & \textbf{14.04} & & 0.376 & \textbf{0.167} & 0.405 & 0.349  & 0.286 & & \textbf{3.22} & \textbf{7.92} & \textbf{7.34}\\
				NDC-substances & 0.272 & - & 0.244 & 0.251  & \textbf{0.202} & & \textbf{33.69} & \textbf{40.13} & \textbf{39.66} & &  0.521  & - & 0.591 & 0.492  & \textbf{0.453} & & \textbf{45.30} & \textbf{55.38} & \textbf{54.99}\\
				tags-ubuntu & 0.091 & \textbf{0.019}  & 0.182  & 0.034 & 0.033 & & \textbf{42.33} & \textbf{43.70} & \textbf{43.55} & & 0.148 & 0.067 & 0.191 & \textbf{0.020} & 0.074 & & \textbf{14.25} & \textbf{15.57} & \textbf{15.43}\\
				tags-math & 0.095 & 0.066 & 0.278 & 0.073 & \textbf{0.011} & & \textbf{42.75} & \textbf{45.60} & \textbf{45.41} & & 0.209 & \textbf{0.053} & 0.286 & 0.113 & 0.079 & & \textbf{21.38} & \textbf{23.12} & \textbf{22.99}\\
				threads-ubuntu & 0.011 & 0.137 & - & \textbf{0.008} & 0.009 & & \textbf{1.28} & \textbf{1.75} & \textbf{1.75} & & \textbf{0.004} & 0.130  & - & \textbf{0.004} & \textbf{0.004} & & -1,346 & -1.72 & -1.72\\
				threads-math & 0.041 & 0.163 & - & \textbf{0.014}  & 0.033 & & \textbf{15.79} & \textbf{16.66} & \textbf{16.52} & & 0.006 & 0.138 & - & \textbf{0.001} & 0.005 & & -1.49 & -0.98 & \textbf{0.96}\\
				coauth-DBLP & 0.224 & - & - & 0.191  & \textbf{0.032} & & \textbf{55.86} & \textbf{74.95} & \textbf{73.45} & & 0.215 & - & - & 0.214 & \textbf{0.192} & & \textbf{2.87} & \textbf{6.73} & \textbf{6.46}\\
				coauth-geology & 0.178 & - & - & 0.157 & \textbf{0.040} & & \textbf{31.13} & \textbf{45.08} & \textbf{44.06} & & 0.086 & - & - & 0.085 & \textbf{0.069} & & -0.10 & \textbf{1.10} & \textbf{0.84}\\
				coauth-history & 0.033 & - & - & 0.030 & \textbf{0.009} & & \textbf{1.74} & \textbf{1.77} & \textbf{1.63} & & \textbf{0.001} & - & - & \textbf{0.001} & \textbf{0.001} & & -0.86 & - & \textbf{0.57}\\
				\hline
				\textbf{Average} & 0.158 & 0.095 & 0.229 & 0.110 & \textbf{0.066} & & & & & & 0.193 & 0.130 & 0.302 & 0.147 & \textbf{0.128} & & & & \\
				\hline
				\multicolumn{7}{l}{-: out of time (taking more than 10 hours) or out of memory} \\
			\end{tabular}
		}
	\end{center}
\end{table*}

\smallsection{Optimization Scheme:}
Having defined the objective, we describe how \methodauto minimizes it. To avoid empty groups, $L\leq \log_2 |V|$ should hold, and the number of levels $L$ is initialized to $\floor{\log_2 |V|}$. 

\blue{Since there are infinitely many combinations of level weights $w_1,\cdots,w_L$, we propose an efficient greedy optimization scheme} described in Algorithm~\ref{alg:optimization}, where some fraction of hyperedges created at a lower level are replaced with those newly created at a higher level, repeatedly, until Eq.~\eqref{eq:objective} converges.

Specifically, \methodauto first generates a hypergraph by \hypercl, which is equivalent to \method with $L=1$ (line~\ref{alg:auto:init}).
This is equivalent to set $w_1$ to $1$ and set $w_\ell$ to $0$ for all $\ell > 1$.
Then at each level $\ell$ from $2$ to $L$, we search for an optimal fraction of hyperedges created at level $\ell-1$ to be replaced with those newly created at level $\ell$ (line~\ref{alg:auto:opt}). 
\blue{Note that only hyperedges of size $\frac{|V|}{2^{\ell-1}}$ or smaller can be replaced.}
If the replacement strictly decreases the hyperedge homogeneity distance, then \methodauto updates the current synthetic hypergraph (line~\ref{alg:auto:update}). This is equivalent to decrease $w_{\ell-1}$ and increase $w_{\ell}$ by the same amount.
Otherwise, we return the current synthetic hypergraph (line~\ref{alg:auto:break}).
We fix the update resolution $p$ to $0.05$ throughout this work. We note that the quality of generated hypergraphs is empirically insensitive to the choices of $p$.

\begin{figure}[!t]
	\vspace{-2mm}
	\centering
	\subfigure[\method (generation)]{
		\includegraphics[width=0.42\columnwidth]{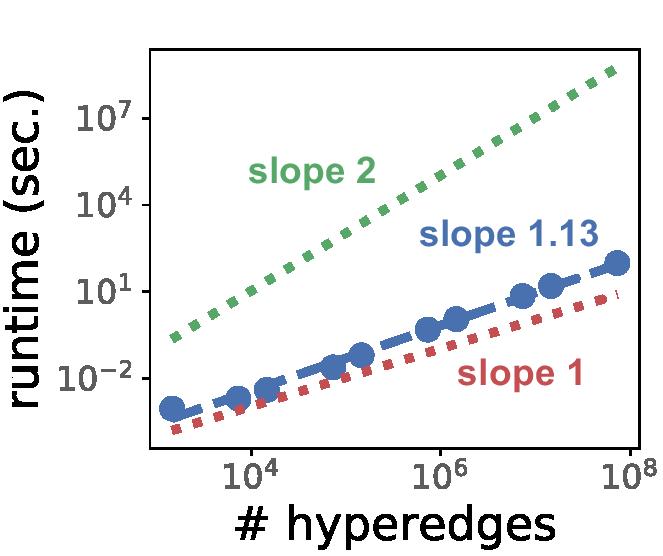}
		\label{fig:nrGroup}
	}
	\hspace{5pt}
	\subfigure[\methodauto (generation \& fitting)]{
		\includegraphics[width=0.42\columnwidth]{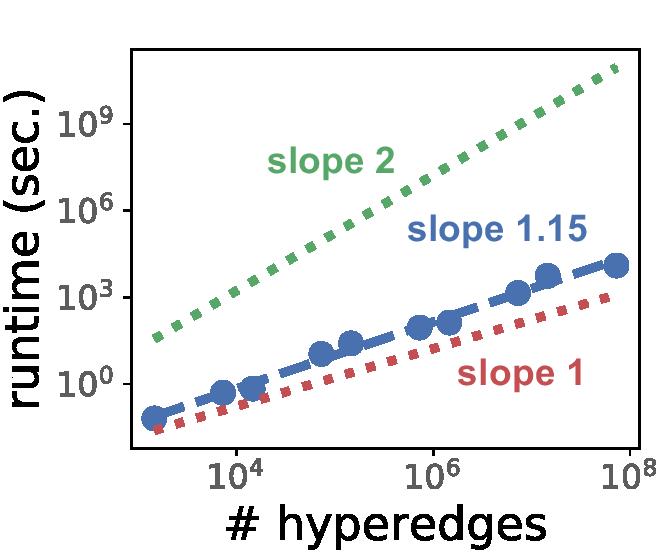}
		\label{fig:overallResult}
	}
	\vspace{-3pt}
	\caption{\method and \methodauto scale near linearly with the size of the considered hypergraph.~\label{fig:scalability}}
\end{figure}

\subsection{Empirical Evaluation of the Quality of Generated Hypergraphs}
\label{sec:method:evaluation}
How well do the hypergraphs generated by \methodauto reproduce the structural properties of the input hypergraphs? We evaluate its effectiveness  by comparing them with four strong baselines: \hypercl, \hyperpa~\cite{do2020structural}, \hyperff~\cite{kook2020evolution}, and na\"ively tuned \method.\footnote{We set the number of levels $L$ same as \methodauto and assign the weights $\{w_1,\cdots,w_{L}\}$ uniformly equal, i.e., $w_i=1/L$ $\forall 1\leq i\leq L$.}
We describe the detailed experimental settings at Appendix~\ref{sec:appendix:experimental_settings}.

To measure the similarity between the distributions derived from the real-world hypergraph and the generated hypergraph, we use the Kolmogorov-Smirnov D-statistic, defined as $D=\max_{x}\{|F'(x)-F(x)|\}$, where $x$ is a value of the considered random variable, and $F'$ and $F$ are the cumulative distribution functions of the real and corresponding generated distributions.

\smallsection{Observations~\ref{obs:density} and ~\ref{obs:overlapness}:}
In Table~\ref{tab:density_overlapness_homogeneity}, we report the D-statistics between the distributions of egonet density and egonet overlapness in real-world hypergraphs and corresponding synthetic hypergraphs. \textbf{\methodauto generates hypergraphs that consist of egonets that are structurally most similar to those in real-world hypergraphs.} 
Specifically, \methodauto gave \textbf{$\mathbf{2.06\times}$ more similar} egonet density distribution and \textbf{$\mathbf{2.35\times}$ more similar} egonet overlapness distribution than recently proposed \hyperpa. 

\smallsection{Observations~\ref{obs:pair} and ~\ref{obs:triple}:} We visually and statistically test whether the hypergraphs generated by \methodauto follow observations~\ref{obs:pair} and \ref{obs:triple}. In Table~\ref{tab:experiment:pair}, we illustrate the distributions of the number of hyperedges overlapping at each pair and each triple of nodes. Compared to \hypercl, \textbf{\methodauto better reproduce\blue{s} the degrees of pairs and triples of nodes.} This is statistically confirmed in Table~\ref{tab:pair_triple}, where \methodauto gives the smallest D-statistic. In addition, these distributions are heavy-tailed in most datasets, as seen from the fact that at least one likelihood ratio is positive
(see Section~\ref{sec:observation:pair_triple_level} for the details of the statistical test).

\smallsection{Observation~\ref{obs:hyperedge_locality}:}
From the results in Table~\ref{tab:density_overlapness_homogeneity}, we can see that \textbf{the D-statistics between the distributions of hyperedge homogeneity in real-world and corresponding hypergraphs generated by \methodauto are extremely small}. Since the objective of \methodauto is to reduce the $HHD$, it naturally reproduce hyperedge homogeneity better than \hypercl, which surprisingly outperforms \method when its parameters are na\"ively set. This result suggests the effectiveness of the proposed optimization scheme. As seen in Table~\ref{loglikelihood-ratio-table-homogeneity}, the distributions of hyperedge homogeneity in hypergraphs generated by \methodauto are heavy-tailed (see Section~\ref{sec:observation:pair_triple_level} for the details of the statistical test).

\subsection{Scalability of \method and \methodauto}
In this subsection, we analyze the scalability of \method and \methodauto both theoretically and experimentally. Noteworthy, we show empirically that both \method and \methodauto scale almost linearly with the size of the considered hypergraph.

In fact, while some baselines are intractable in particular datasets, \method and \methodauto are scalable enough to be executed in all considered datasets. 
The scalability of \hyperpa heavily depends on the sizes of hyperedges, and thus does not work in hypergraphs that includes large-sized hyperedges (i.e., email-Eu, NDC-substances, coauth-DBLP, coauth-geology, and coauth-history).
\hyperff depends on the number of nodes, and does not work in large datasets with many nodes (i.e., threads-ubuntu, threads-math, coauth-DBLP, coauth-geology, and coauth-history).

Given the number of levels and weights of each level, how much time does it take to run \method? Assume that all sets and maps are implemented using hash tables. For each hyperedge $e$, level $\ell$ and group $g$ are selected in $O(1)$ time. In addition, since each node is sampled independently, $|e|$ nodes are selected in $O(|e|\cdot (1+\epsilon))$ time, where $\epsilon$ is due to the possibility of collisions (i.e., nodes selected multiple times for a hyperedge). 
The term $\epsilon$ depends on the degrees of nodes and the sizes of hyperedges.
We note that $\epsilon$ is empirically very small in the considered datasets.
Hence, generating $|E|$ hyperedges takes $O(\sum_{e\in E} (|e|\cdot (1+\epsilon)))$ time.  
In \methodauto, we consider the replacement step. At each level, at most $\frac{1}{p}\cdot |E|=O(|E|)$ hyperedges are (temporarily) replaced, taking $O(\sum_{e\in E}(|e|\cdot (1+\epsilon)))$ time. Since the maximum number of levels is $\log_2 |V|$, \methodauto takes  $O(\log_2 |V| \cdot \sum_{e\in E}(|e|\cdot (1 + \epsilon)))$ time in total.

In Figure~\ref{fig:scalability}, we measure the runtimes of \method and \methodauto with synthetic hypergraphs of different sizes.
They are generated by upscaling the smallest hypergraph, email-Enron by $5$ to $50,000$ times, using \method. Both \method and \methodauto scale almost linearly with the size of the considered hypergraph. Specifically, \methodauto generates and fits a synthetic hypergraph with 0.7 billion hyperedges within few hours. 
We describe the detailed experimental settings at Appendix~\ref{sec:appendix:experimental_settings}.

%% file: 070summary.tex
In this work, we investigate the structural properties regarding the overlaps of hyperedges of thirteen real-world hypergraphs from six domains.
To this end, we define several principled measures, and based on the observations, we develop a realistic hypergraph generative model.
We summarize our contributions as follows.

\setlength{\leftmargini}{1em}
\begin{itemize}
	\item \textbf{Observations in Real-world Hypergraphs:} We discover three unique properties of the overlaps of hyperedges in real-world hypergraphs. We verify these properties using randomized hypergraphs where both the degrees of nodes and the sizes of hyperedges are well preserved.
	\item \textbf{Novel Measures:} We propose the overlapness and homogeneity of hyperedges. We demonstrate through an axiomatic approach that overlapness is a principled measure. Homogeneity reveals an interesting overlapping pattern, based on which we develop a realistic generative model.
	\item \textbf{Realistic Generative Model:} We propose \method, a hypergraph generative model that accurately reproduces the overlapping patterns of hyperedges in real-world hypergraphs. We also provide \methodauto, which automatically fits the parameters of \method to a given graph. They generate and fit a hypergraph with $0.7$ billion hyperedges within few hours.
\end{itemize}

\noindent \textbf{Reproducibility:}  The source code and datasets used in this work are available at \blue{\url{https://github.com/young917/www21-hyperlap}}.

%% file: 080appendix.tex


\section{Appendix: Axioms of Overlapness~\label{sec:appendix:overlapness_axiom}}
We systematically analyze overlapness defined in Section~\ref{sec:observation:egonet_level} by comparing with possible baselines and proving that the metric satisfies all the proposed axioms.

\smallsection{Baselines:}
Due to the simplicity and intuitiveness of the aforementioned axioms, one might hypothesize that it is trivial to satisfy them. However, as seen in Table~\ref{tab:axioms}, none of the other possible baseline metrics obey all three axioms. We consider five different baseline metrics including two basic set operations: 

\begin{itemize}
    \setlength\itemsep{0.2em}
    \item{\makebox[3.5cm]{\textbf{Intersection:}\hfill}$|\bigcap_{e\in \mathcal{E}}e|$}
    \item{\makebox[3.5cm]{\textbf{Union Inverse:}\hfill}$1/|\bigcup_{e\in \mathcal{E}}e|$}
    \item{\makebox[3.5cm]{\textbf{Jaccard Index:}\hfill}$|\bigcap_{e\in \mathcal{E}}e|/|\bigcup_{e\in \mathcal{E}}e|$}
    \item{\makebox[3.5cm]{\textbf{Overlap Coefficient:}\hfill}$|\bigcap_{e\in \mathcal{E}}e|/\min_{e\in \mathcal{E}}|e|$}
    \item{\makebox[3.5cm]{\textbf{Density~\cite{hu2017maintaining}:}\hfill}$|\mathcal{E}|/|\bigcup_{e\in \mathcal{E}}e|$}
\end{itemize}

\noindent Using the intersection of multiple hyperedges as the measure is applicable to only a small number of hyperedges (i.e., small $k$) due to its strict condition that nodes should be included in all the given hyperedges. Accordingly, other possible measures to gauge the overlaps such as the Jaccard index or the overlap coefficient, which use intersection size as a numerator, face the same challenge. The inverse of the union meets Axiom~\ref{axm:num_nodes}, while it does not satisfy Axiom~\ref{axm:num_hyperedges} and Axiom~\ref{axm:size_hyperedges}. The density of the hyperedges satisfies Axiom~\ref{axm:num_hyperedges} and Axiom~\ref{axm:num_nodes}, while it does not satisfy Axiom~\ref{axm:size_hyperedges}, which is clear from the example discussed in Section~\ref{sec:observation:egonet_level}. We provide detailed examples and reasons why each baseline measure does not satisfy at least one axiom in the online appendix~\cite{online2020appendix}.

\smallsection{Proof of Theorem~\ref{thm:overlapness}:}
We show that overlapness meets all three axioms discussed in Section~\ref{sec:observation:egonet_level}.
That is, we prove Theorem~\ref{thm:overlapness} by proving Lemmas~\ref{lemma:axiom:one}, \ref{lemma:axiom:two}, and \ref{lemma:axiom:three},  which Theorem~\ref{thm:overlapness} follows from.

\begin{lma} \label{lemma:axiom:one}
    Overlapness meets Axiom~\ref{axm:num_hyperedges}.
\end{lma}
\textsc{\textbf{Proof.}}
Considering the conditions in Axiom~\ref{axm:num_hyperedges}, we compare the overlapness of $\mathcal{E}$ and $\mathcal{E}'$:

\vspace{-3mm}
\begin{align*}
   o(\mathcal{E}') - o(\mathcal{E}) = \frac{\sum_{e'\in \mathcal{E}'}|e'|}{|\bigcup\nolimits_{e'\in \mathcal{E}'}e'|} - \frac{\sum_{e\in \mathcal{E}}|e|}{|\bigcup\nolimits_{e\in \mathcal{E}}e|} = \frac{n\cdot (|\mathcal{E}'|-|\mathcal{E}|)}{|\bigcup\nolimits_{e\in \mathcal{E}}e|}
\end{align*}
from the conditions $|\bigcup\nolimits_{e\in\mathcal{E}}e|=|\bigcup\nolimits_{e'\in\mathcal{E}'}e'|$ and $|e|=|e'|=n,\ \forall e\in\mathcal{E}, \ \forall e'\in\mathcal{E}'$. Since the number of hyperedges in $\mathcal{E}'$ is larger than that in $\mathcal{E}$ (i.e., $|\mathcal{E}'|>|\mathcal{E}|$), $o(\mathcal{E}')>o(\mathcal{E})$ holds. This implies Axiom~\ref{axm:num_hyperedges}. \hfill $\blacksquare$

\begin{lma} \label{lemma:axiom:two}
    Overlapness meets Axiom~\ref{axm:num_nodes}.
\end{lma}
\textsc{\textbf{Proof.}}
Considering the conditions in Axiom~\ref{axm:num_nodes}, we compare the overlapness of $\mathcal{E}$ and $\mathcal{E}'$:
\begin{align*}
    \frac{o(\mathcal{E}')}{o(\mathcal{E})} = 
    \frac{\sum_{e'\in \mathcal{E}'}|e'|}{|\bigcup\nolimits_{e'\in \mathcal{E}'}e'|} \bigg/ \frac{\sum_{e\in \mathcal{E}}|e|}{|\bigcup\nolimits_{e\in \mathcal{E}}e|} = \frac{|\bigcup\nolimits_{e\in \mathcal{E}}e|}{|\bigcup\nolimits_{e'\in \mathcal{E}'}e'|}
\end{align*}
from the conditions $|\mathcal{E}|=|\mathcal{E'}|=n$ and $|e_i|=|e'_i|,\ \forall i\in \{1,...,n\}$. Since the number of nodes in $\mathcal{E}$ is more than $\mathcal{E}'$ (i.e., $|\bigcup\nolimits_{e\in \mathcal{E}}e|>|\bigcup\nolimits_{e' \in \mathcal{E}'}e'|$, $\frac{o(\mathcal{E}')}{o(\mathcal{E})}>1$ holds, and thus $o(\mathcal{E}')>o(\mathcal{E})$. This implies Axiom~\ref{axm:num_nodes}. \hfill $\blacksquare$

\begin{lma} \label{lemma:axiom:three}
    Overlapness meets Axiom~\ref{axm:size_hyperedges}.
\end{lma}
\textsc{\textbf{Proof.}}
Considering the conditions in Axiom~\ref{axm:size_hyperedges}, we compare the overlapness of $\mathcal{E}$ and $\mathcal{E}'$:
\begin{align*}
    o(\mathcal{E}') - o(\mathcal{E}) = \frac{\sum_{e'\in \mathcal{E}'}|e'|}{|\bigcup\nolimits_{e'\in \mathcal{E}'}e'|} - \frac{\sum_{e\in \mathcal{E}}|e|}{|\bigcup\nolimits_{e\in \mathcal{E}}e|} =
    \frac{\sum\nolimits_{k=1}^{n}(|e'_k|-|e_k|)}{|\bigcup\nolimits_{e\in\mathcal{E}}e|}
\end{align*}
from the conditions $|\mathcal{E}|=|\mathcal{E}'|=n$ and $|\bigcup\nolimits_{e\in\mathcal{E}}e|=|\bigcup\nolimits_{e'\in\mathcal{E}'}e'|$. Since $|e_i|<|e'_i|$, and $|e_j|\leq|e'_j|,\ \forall j\in \{1,...,n\} \setminus \{i\}$, $o(\mathcal{E}')>o(\mathcal{E})$ holds. This implies Axiom~\ref{axm:size_hyperedges}.  \hfill $\blacksquare$

\section{Appendix: Experimental Settings}\label{sec:appendix:experimental_settings}
We describe the environmental settings where we conducted experiments covered in this paper.

\smallsection{Machines:} 
We conducted all the experiments on a machine with an AMD Ryzen 9 3900X CPU and 128GB RAM.

\smallsection{Datasets:} 
We used thirteen real-world hypergraphs from six different domains. See Section~\ref{sec:prelim:datasets} for details of the datasets.

\smallsection{Baselines:}
We evaluate \method and \methodauto by comparing with following three baseline models:

\setlength{\leftmargini}{1em}
\begin{itemize}
    \item \textbf{\hypercl:} This model, which is described in Section~\ref{sec:prelim:random_hypergraphs}, is a generalization of the FCL model to hypergraphs. It preserves well the degree distribution of the input hypergraph.
    \item \textbf{\hyperpa~\cite{do2020structural}:} This model, which is described in Section~\ref{sec:related}, extends the preferential attachment model to hypergraphs so that each new node forms a hyperedge with each subset of nodes, rather than individual nodes, with probability proportional to the number of the hyperedges containing the subset. 
    \item \textbf{\hyperff~\cite{kook2020evolution}:} This model, which is described in Section~\ref{sec:related}, extends the forest fire model to hypergraphs. The model has two parameters, which are the burning and expanding rates. We set them to $0.51$ and $0.2$, as suggested in the paper.
\end{itemize}

\smallsection{Implementations:} 
We implemented \hypercl and \method using C++. For \hyperpa and \hyperff, we used their open-source implementations in Python. \footnote{The open-source implementations are available at  \url {https://github.com/manhtuando97/KDD-20-Hypergraph} and \url{https://github.com/yunbum-kook/icdm20-hyperff}.}